\documentclass[useAMS,usenatbib,usegraphicx]{mn2e}
\usepackage{psfig}   
\usepackage{graphicx}
\usepackage{amssymb}
\usepackage{subfigure}

\title[Dynamical histories of IC\,348 and NGC\,1333]{Dynamical histories of the IC\,348 and NGC\,1333 star-forming regions in Perseus}

\author[R.~J.~Parker \& C.~Alves~de~Oliveira]{
  Richard J.~Parker$^{1}$\thanks{E-mail: R.Parker@sheffield.ac.uk}\thanks{Royal Society Dorothy Hodgkin Fellow} and Catarina Alves de Oliveira$^2$ \vspace*{0.1cm}\\
   $^1$ Department of Physics and Astronomy, The University of Sheffield, Hicks Building, Hounsfield Road, Sheffield, S3 7RH, UK \\
   $^2$ European Space Agency, c/o STScI, 3700 San Martin Drive, Baltimore, MD 21218, USA}

\begin{document}

                             
\pagerange{\pageref{firstpage}--\pageref{lastpage}} \pubyear{2017}

\maketitle

\label{firstpage}

\begin{abstract}
We present analyses of the spatial distributions of stars in the young (1 -- 3\,Myr) star-forming regions IC\,348 and NGC\,1333 in the Perseus Giant Molecular Cloud. We quantify the spatial structure using the $\mathcal{Q}$-parameter and find that both IC\,348 and NGC\,1333 are smooth and centrally concentrated with $\mathcal{Q}$-parameters of 0.98 and 0.89 respectively. Neither region exhibits mass segregation ($\Lambda_{\rm MSR} = 1.1^{+0.2}_{-0.3}$ for IC\,348 and $\Lambda_{\rm MSR} = 1.2^{+0.4}_{-0.3}$ for NGC\,1333, where $\Lambda_{\rm MSR} \sim 1$ corresponds to no mass segregation), nor do the most massive stars reside in areas of enhanced stellar surface density compared to the average surface density, according to the $\Sigma_{\rm LDR}$ method.

We then constrain the dynamical histories and hence initial conditions of both regions by comparing the observed values to $N$-body simulations at appropriate ages. Stars in both regions likely formed with sub-virial velocities which contributed to merging of substructure and the formation of smooth clusters. The initial stellar densities were no higher than $\rho \sim 100 - 500$\,M$_\odot$\,pc$^{-3}$ for IC\,348 and $\rho \sim 500 - 2000$\,M$_\odot$\,pc$^{-3}$ for NGC\,1333. These initial densities, in particular that of NGC\,1333, are high enough to facilitate dynamical interactions which would likely affect $\sim$10\,per cent of protoplanetary discs and binary stars.
\end{abstract}


\section{Introduction}

Most stars form in regions where the local stellar density exceeds that of the Galactic field by several orders of magnitude \citep{Porras03,Lada03,Bressert10}. However, it is currently unclear what fraction of stars are either born in, or experience, a dense stellar environment before they disperse into the field. If most star-forming regions are initially very dense (e.g.\,\,$>$100\,stars\,pc$^{-3}$), then dynamical encounters can alter or disrupt the orbits of binary stars \citep[e.g.][]{Kroupa95a,Marks11,Parker12b} and planetary systems \citep[e.g.][]{Bonnell01b,Adams06,Parker12a}. It is therefore crucial to determine the typical maximum density of star-forming regions before they dissolve into the Galactic disc. 

An initially dense star-forming region will expand due to two-body relaxation \citep{Gieles12,Moeckel12,Parker12d} and so the observed present-day density may be much lower than the initial density. This degeneracy can be overcome by incorporating information on the spatial structure of the region. Stars form in filaments \citep{Andre14}, and this results in a hierarchically substructured spatial distribution for the stars in some observed regions \citep{Cartwright04,Sanchez09}, and in hydrodynamical simulations of star formation \citep{Bonnell03,Bate09,Dale12a,Dale14}. 

In previous work, we have shown that combining the spatial structure with measures of the relative spatial distribution of the most massive stars, compared to low-mass stars, places constraints on the initial density and virial state of a region \citep{Parker14b,Parker14e,Wright14}. Two-body relaxation erases primordial substructure and leads to the most massive stars residing in areas of relatively high local density \citep{Parker14b}, and violent relaxation can facilitate dynamical mass segregation on timescales commensurate with the ages of young star clusters \citep{McMillan07,Allison10}. 

In order to determine the initial or maximum density of a star-forming region, a reasonably complete membership census is required, as well as mass determinations for all objects. Thus far, only four star-forming regions -- the Orion Nebula Cluster, Taurus, $\rho$~Ophiuchus and Cyg~OB2 -- have had their initial densities constrained by this method. In this paper, we use the most recent censuses of the nearby ($<$350\,pc), young (1 -- 3\,Myr) IC\,348 and NGC\,1333 star-forming regions in the Perseus Giant Molecular Cloud from \citet{Luhman16} to determine the spatial structure and relative spatial distributions of the most massive stars compared to other stars and brown dwarfs. 

The paper is organised as follows. In Section~\ref{data} we describe the datasets used, in Section~\ref{method} we outline the methods used to quantify the spatial distributions and in Section~\ref{results} we present the results for IC\,348 and NGC\,1333. We compare these results to $N$-body simulations of the evolution of star clusters to constrain the initial density of both regions in Section~\ref{nbody_comp}. We provide a discussion in Section~\ref{discuss} and we conclude in Section~\ref{conclude}.

\section{Datasets}
\label{data}

For the observational sample, we adopted the most recent census of IC~348 and NGC~1333 clusters presented in \citet{Luhman16}, where tens of new low-mass stars and brown dwarf candidate members have been confirmed spectroscopically. Additionally, \citet{Luhman16} have classified spectroscopically several previously known members, resulting in an unprecedented census with a number of members of 478 for IC~348 and 203 for NGC~1333. In both clusters, all members have measured magnitudes and an estimate of extinction, as well as a spectral type, except for 19 members of IC 348 and 41 members of NGC 1333.   

To estimate the masses of the members of the clusters, we first converted spectral types to temperatures, adopting the temperature scale from \citet{Schmidt82} for stars earlier than M0, and the scale from  \citet{Luhman03b} for sources with spectral type between M0 and M9.5. For the L dwarfs, we applied the scale proposed by \citet{Lodieu08}. The bolometric luminosity was calculated by de-reddening the magnitudes of the targets with the extinction law from \citet{Rieke85}, using a distance to the clusters of 300~pc for IC~348 and 235~pc for NGC~1333, and bolometric corrections from \citet{Pecaut13}, \citet{Herczeg15}, and \citet{Dahn02}. Masses were derived from the evolutionary models \citep{Baraffe98,Chabrier00,Siess00} according to each target's location on the HR diagram. For the lowest mass brown dwarfs (spectral types later than $\sim$M8), luminosities become more uncertain both in the data and the models, and therefore masses are derived only from temperatures using the models from \citet{Chabrier00}.

\begin{figure}
\begin{center}
\rotatebox{270}{\includegraphics[scale=0.35]{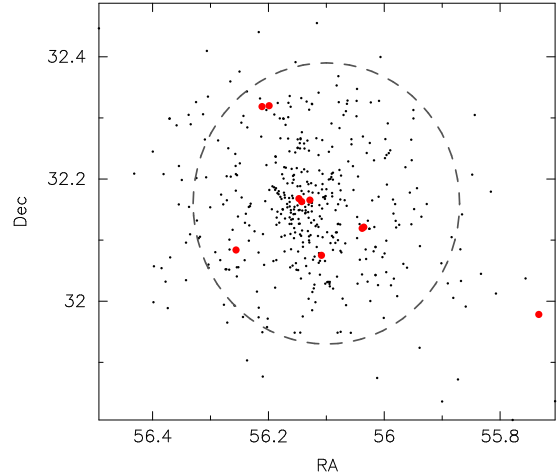}}
\end{center}
\caption[bf]{Map of objects in IC348. The red points indicate the 10 most massive stars. The area within the dashed line is the 14' radius from the centre which is observationally complete \citep{Luhman16}.}
\label{ic348_map}
\end{figure}

\begin{figure}
\begin{center}
\rotatebox{270}{\includegraphics[scale=0.35]{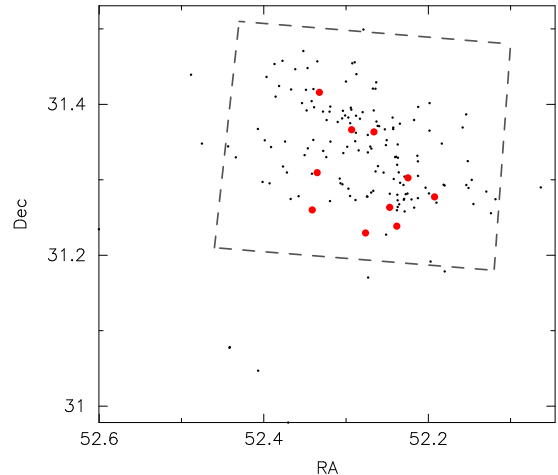}}
\end{center}
\caption[bf]{Map of objects in NGC1333. The red points indicate the 10 most massive stars. The area within the dashed lines is the ACIS-I field from \citet{Luhman16}, which is observationally complete.}
\label{ngc1333_map}
\end{figure}

In Fig.~\ref{ic348_map} we show the positions of the objects in IC\,348, and in Fig.~\ref{ngc1333_map} we show the positions of the objects in NGC\,1333. In both plots the ten most massive stars in each region are shown by the large red points. Both regions have been extensively surveyed, and we show the areas that are observationally complete by the dashed lines. In IC\,348 this corresponds to an area with radius 14' from the central star BD +31 643 and in NGC\,1333 the area covered by the Chandra X-ray Observatory Advanced CCD Imaging Spectrometer (ACIS-I). In addition to the full datasets, we will also perform our analyses only on objects within these areas.


\section{Methods}
\label{method}

In this section we outline the techniques used to quantify the spatial distributions of stars and brown dwarfs in IC\,348 and NGC\,1333.

\subsection{Spatial structure}

We determine the amount of structure in a star forming region by measuring the $\mathcal{Q}$-parameter. The $\mathcal{Q}$-parameter was pioneered by \citet{Cartwright04,Bastian09,Cartwright09a,Cartwright09b} and combines the normalised mean edge length of the minimum spanning tree (MST) of all the stars in the 
region, $\bar{m}$, with the normalised correlation length between all stars in the region, 
$\bar{s}$. The level of substructure is determined by the following equation:
\begin{equation}
\mathcal{Q} = \frac{\bar{m}}{\bar{s}}.
\end{equation}
A substructured association or region has $\mathcal{Q}<0.8$, whereas a smooth, centrally concentrated cluster has $\mathcal{Q}>0.8$. The $\mathcal{Q}$-parameter has the advantage of being independent of the density of the star forming region, and purely measures the level of substructure present. 

\subsection{Mass segregation}

In order to quantify the amount of mass segregation present in the clusters, we use the $\Lambda_{\rm MSR}$ method, introduced by \citet{Allison09a}.  This 
constructs a minimum spanning tree (MST) between a chosen subset of stars and then compares this MST to the average MST length of many 
random subsets. 

We find the MST of the $N_{\rm MST}$ stars in the chosen subset and
compare this to the MST of sets of $N_{\rm MST}$ random  stars in the
region. If the length of the MST of the chosen subset is shorter than
the average length of the MSTs for the  random stars then the subset
has a more concentrated distribution and is said to be mass segregated. Conversely, if the MST  length of the chosen subset is
longer than the average MST length, then the subset has a less
concentrated distribution, and is  said to be inversely mass
segregated \citep[see e.g.][]{Parker11b}. Alternatively, if the MST length of the chosen subset is
equal to the random MST length,  we can conclude that no mass
segregation is present.

By taking the ratio of the average (mean) random MST length to the subset MST
length, a quantitative measure of the degree of  mass segregation
(normal or inverse) can be obtained. We first determine the subset MST
length, $l_{\rm subset}$. We then  determine the average length of
sets of $N_{\rm MST}$ random stars each time, $\langle l_{\rm average}
\rangle$. There is a dispersion  associated with the average length of
random MSTs, which is roughly Gaussian and can be quantified as the
standard deviation  of the lengths  $\langle l_{\rm average} \rangle
\pm \sigma_{\rm average}$. However, we conservatively estimate the lower (upper) uncertainty 
as the MST length which lies 1/6 (5/6) of the way through an ordered list of all the random lengths (corresponding to a 66 per cent deviation from 
the median value, $\langle l_{\rm average} \rangle$). This determination 
prevents a single outlying object from heavily influencing the uncertainty. 
We can now define the `mass  segregation ratio' 
($\Lambda_{\rm MSR}$) as the ratio between the average random MST pathlength 
and that of a chosen subset, or mass range of objects:
\begin{equation}
\Lambda_{\rm MSR} = {\frac{\langle l_{\rm average} \rangle}{l_{\rm subset}}} ^{+ {\sigma_{\rm 5/6}}/{l_{\rm subset}}}_{- {\sigma_{\rm 1/6}}/{l_{\rm subset}}}.
\end{equation}
A $\Lambda_{\rm MSR}$ of $\sim$ 1 shows that the stars in the chosen
subset are distributed in the same way as all the other  stars,
whereas $\Lambda_{\rm MSR} > 1$ indicates mass segregation and
$\Lambda_{\rm MSR} < 1$ indicates inverse mass segregation,
i.e.\,\,the chosen subset is more sparsely distributed than the other stars.

There are several subtle variations of $\Lambda_{\rm MSR}$. \citet*{Olczak11} propose using the geometric mean to reduce the spread in uncertainties, 
and \citet{Maschberger11} and \citet{Delgado13} propose using the median MST length to reduce the effects of outliers from influencing the results. However, in the subsequent 
analysis we will adopt the original $\Lambda_{\rm MSR}$  from Allison. 

\subsection{Relative local density} 
 
\citet{Maschberger11} and \citet{Kupper11} quantified the relative local density of a subset of objects with respect to the average local stellar density in the cluster. We calculate the local stellar surface density following the prescription of \citet{Casertano85}, modified to account for the analysis in projection. For an individual star the local stellar surface density is given by
\begin{equation}
\Sigma = \frac{N - 1} {\pi r_{N}^2},
\end{equation}
where $r_{N}$ is the distance to the $N^{\rm th}$ nearest neighbouring star (we adopt $N = 10$ throughout this work).

To determine whether a particular subset of stars, $\tilde{\Sigma}_\mathrm{subset}$, reside in areas of significantly higher than average density than the cluster average $\tilde{\Sigma}_\mathrm{all}$, we divide $\tilde{\Sigma}_\mathrm{subset}$ by $\tilde{\Sigma}_\mathrm{all}$ to define a `local density ratio', $\Sigma_{\rm LDR}$:
\begin{equation}
\Sigma_{\rm LDR} = \frac{\tilde{\Sigma}_\mathrm{subset}}{\tilde{\Sigma}_\mathrm{all}}
\end{equation} 
The significance of this measure of the local density of a subset of stars compared to the cluster will be defined by a Kolmogorov-Smirnov (KS) test between the $\Sigma$ values of the subset against the $\Sigma$ values of the rest. 

\citet{Parker14b} and \citet{Parker14e} show that this can then be used to trace the initial density of a star-forming region because in very dense 
($>$1000\,stars\,pc$^{-3}$) regions, the most massive stars act as potential wells and collect retinues of low-mass stars, leading to  $\Sigma_{\rm LDR} >> 1$.

\section{Results}
\label{results}

\subsection{IC\,348}

\begin{figure*}
  \begin{center}
\setlength{\subfigcapskip}{10pt}
\hspace*{-1.5cm}\subfigure[$\Lambda_{\rm MSR}$, high mass]{\label{ic348_data-a}\rotatebox{270}{\includegraphics[scale=0.25]{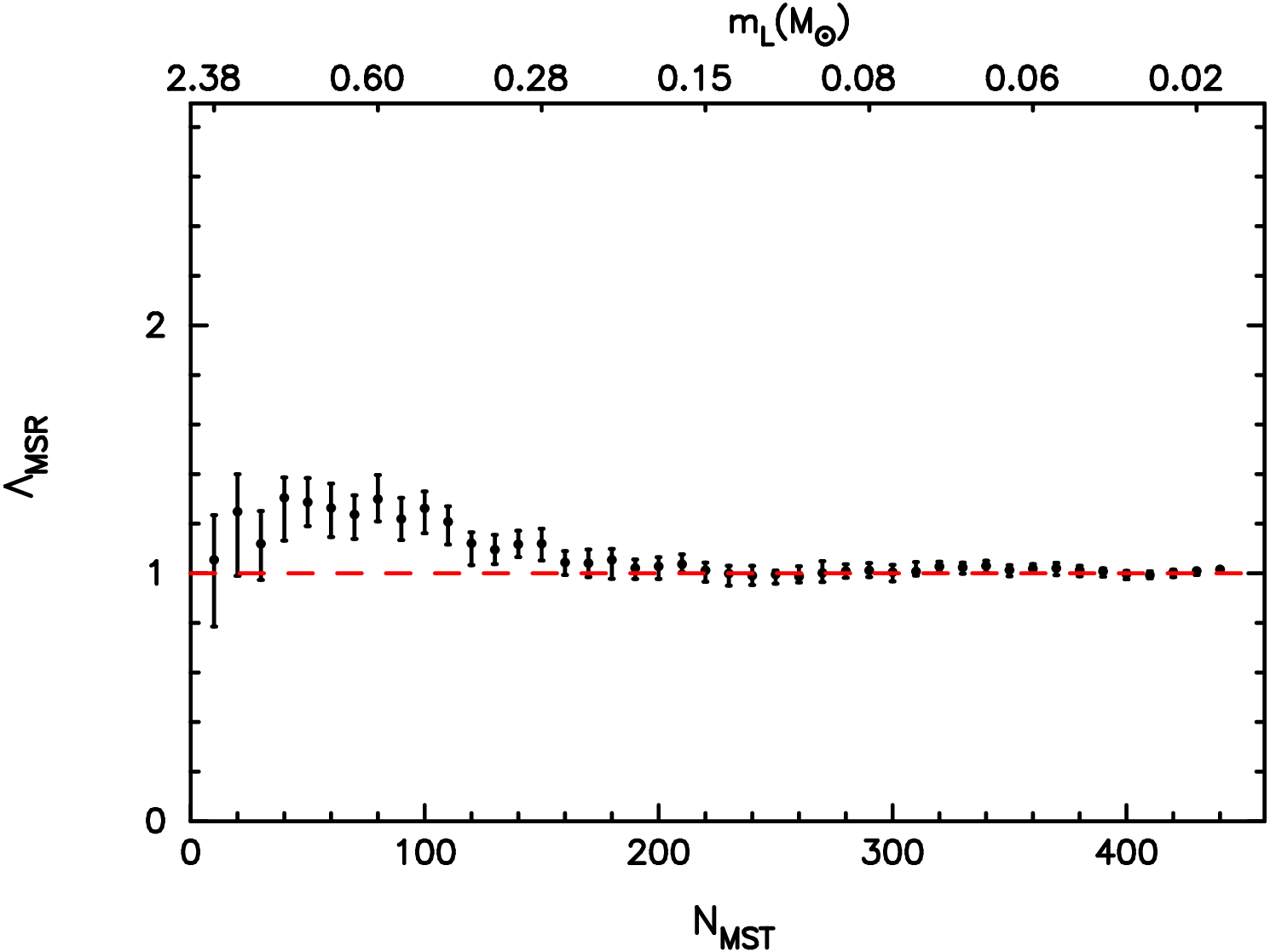}}}
\hspace*{0.3cm} 
\subfigure[$\Lambda_{\rm MSR}$, BDs]{\label{ic348_data-b}\rotatebox{270}{\includegraphics[scale=0.24]{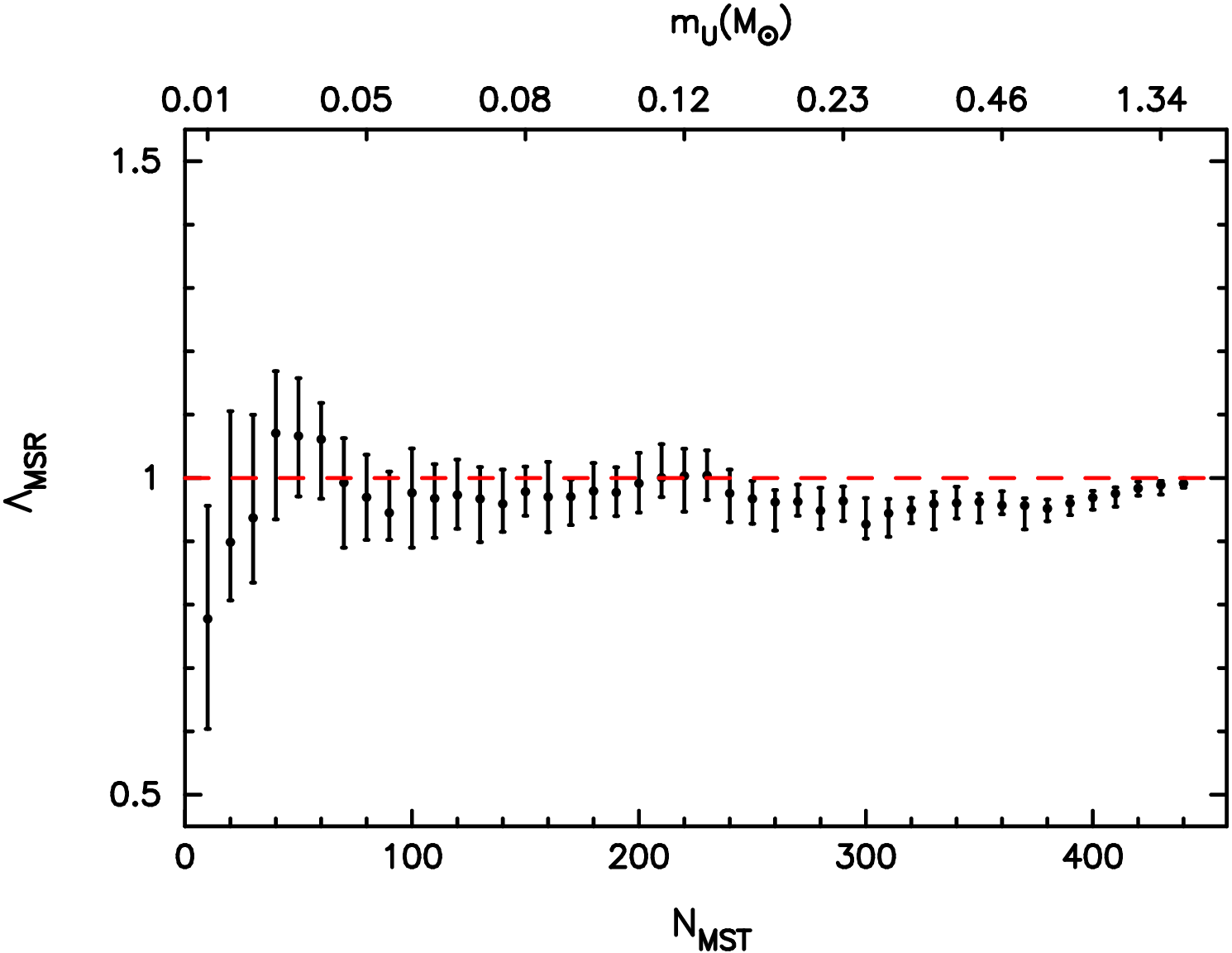}}} 
\hspace*{0.3cm}\subfigure[$\Sigma - m$]{\label{ic348_data-c}\rotatebox{270}{\includegraphics[scale=0.25]{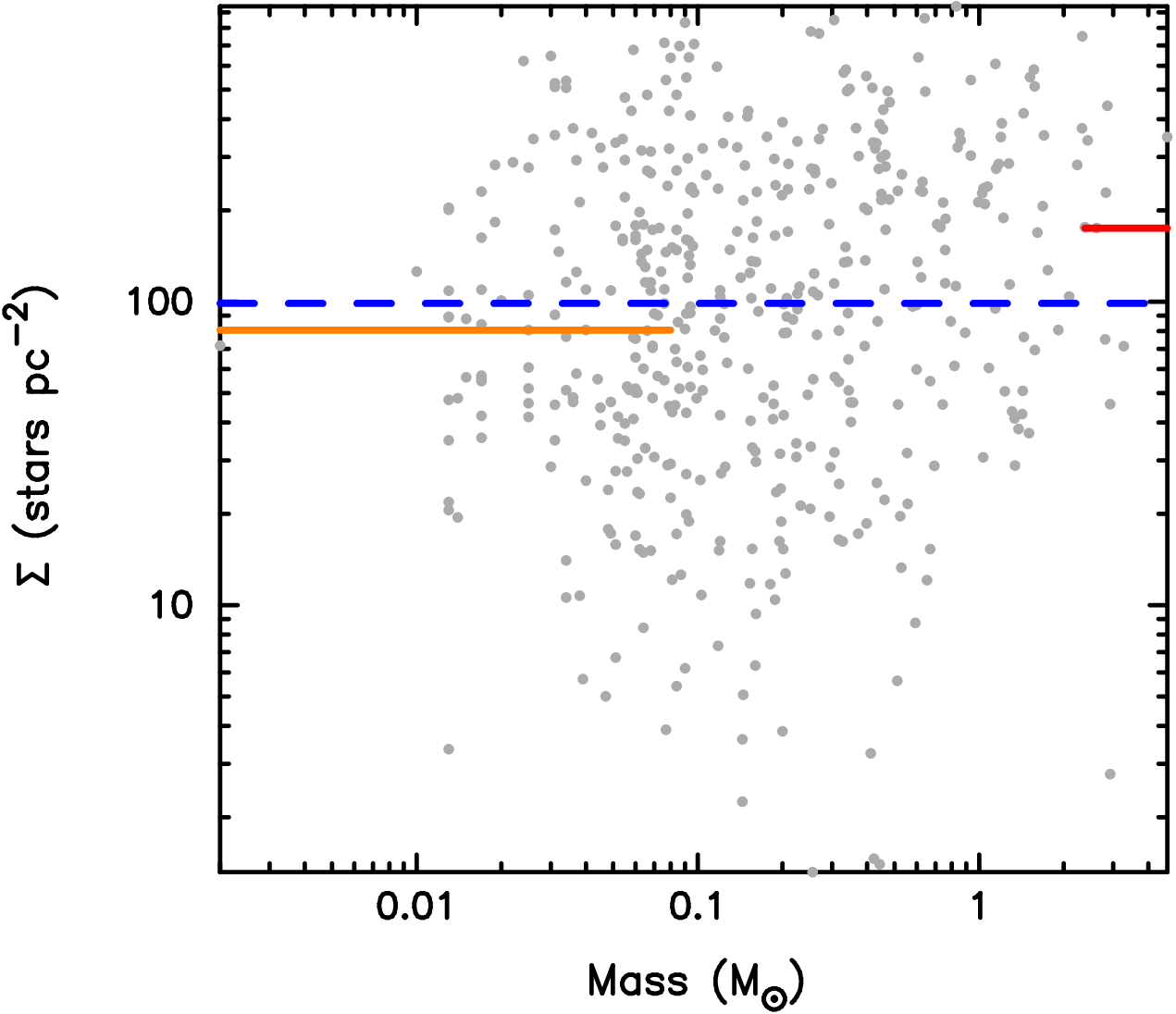}}}
\caption[bf]{Data for IC348. Panel (a) shows $\Lambda_{\rm MSR}$ as a function of the $N_{\rm MST}$ most massive stars. $\Lambda_{\rm MSR} = 1.1$, which is probably not a significant deviation from unity. Panel (b) shows $\Lambda_{\rm MSR}$ as a function of the  $N_{\rm MST}$ least massive objects. Again, the deviation from unity is not likely to be significant. Panel (c) shows the median surface density for the most massive stars (red line), cluster average (blue line) least massive BDs (orange line). Deviations from the median are not significant (KS test p-values of 0.49 and 0.20, respectively). The $\mathcal{Q}$-parameter is 0.98 (centrally concentrated).}
\label{ic348_data}
  \end{center}
\end{figure*}

\subsubsection{Spatial structure}  

We apply the $\mathcal{Q}$-parameter to the IC\,348 dataset and obtain $\mathcal{Q} = 0.98$, which indicates a centrally concentrated distribution.  \citet{Cartwright09b} and \citet{Lomax11} show that centrally concentrated morphologies can be sub-categorised by plotting the distribution of $\bar{m}$, the mean MST length, and $\bar{s}$, the mean separation length, although for complex morphologies this is highly non-trivial \citep{Parker15c}. In our data $\bar{m} = 0.36$ and $\bar{s} = 0.36$, which suggests a smooth, centrally concentrated density profile of the form $N \propto r^{-2.5}$. This is not as concentrated as a \citet{Plummer11} or \citet{King66} profile, suggesting a dynamically young cluster -- we will examine this in more detail in the following Section~\ref{nbody_comp}. 

Interestingly, \citet{Cartwright04} find $\mathcal{Q} = 0.98$ for IC\,348, with $\bar{m} = 0.48$ and $\bar{s} = 0.49$ for 288 objects from the \citet{Luhman03} sample, which is also consistent with an  $N \propto r^{-2.5}$ profile.

If we restrict our determination of the $\mathcal{Q}$-parameter to within the 14' completeness radius from \citet{Luhman16}, we find $\mathcal{Q} = 0.85$, with $\bar{m} = 0.53$ and $\bar{s} = 0.63$. This is a less centrally concentrated spatial profile ($N \propto r^{-1.5}$) than is measured for the full dataset; this is unsurprising as we are pruning the more distant objects from our determination of $\mathcal{Q}$.

\subsubsection{Mass segregation ratio, $\Lambda_{\rm MSR}$}

In Fig.~\ref{ic348_data-a} we show the evolution of $\Lambda_{\rm MSR}$ as a function of the $N_{\rm MST}$ most massive stars in the subset, which is compared to the MSTs of objects randomly drawn from the spatial distribution shown in Fig.~\ref{ic348_map}. $\Lambda_{\rm MSR} = 1.1^{+0.2}_{-0.3}$ for the $N_{\rm MST} = 10$ most massive stars, and does not deviate from unity if we consider fewer (e.g.\,\,$N_{\rm MST} = 4$) or more stars. 

We then show the $\Lambda_{\rm MSR}$ ratio for the $N_{\rm MST}$ least massive objects (brown dwarfs) in Fig.~\ref{ic348_data-b}.  $\Lambda_{\rm MSR} = 0.8^{+0.3}_{-0.1}$ for the $N_{\rm MST} = 10$ least massive brown dwarfs and this plot suggests that these objects are not more or less spatially concentrated than the stars in IC\,348. 

We measure  $\Lambda_{\rm MSR} \sim 1$ for the 14' completeness field, suggesting that our results are not influenced by the inclusion or otherwise of outlying objects.

\subsubsection{Local surface density ratio, $\Sigma_{\rm LDR}$}

The local surface density as a function of mass is shown for all of the stars in our sample in Fig.~\ref{ic348_data-c}. The median surface density for the whole star-forming region is shown by the blue dashed line, the median for the ten most massive stars is shown by the red line and the median surface density of the brown dwarfs is shown by the orange line.  The surface density ratio for the most massive stars is $\Sigma_{\rm LDR} = 1.77$, but a Kolmogorov-Smirnov (KS) test between the two populations gives $D = 0.25$ and a $p-$value of 0.49 that the two populations share the same underlying parent distribution. Based on this, we cannot conclude that the most massive stars reside in areas of higher than average surface density. 

If we compare the surface densities of brown dwarfs to the surface density of the whole star-forming region we have $\Sigma_{\rm LDR} = 0.82$, which suggests the brown dwarfs reside in areas of lower than average surface density. However, the KS test between the two distributions has $D = 0.33$ and the $p-$value is 0.20, again suggesting that the two populations share the same underlying distribution. 

Previous studies have suggested that the most massive objects in IC\,348 have significantly higher surface densities than the lowest mass objects \citep{Kirk12}, and this gradient is also visually apparent in our dataset. The conclusion in \citet{Kirk12} was drawn from analysing a sample of 232 objects, which was split into two bins (high and low mass). This may introduce an inherent bias if one of the subsets in question (e.g.\,\,the most massive stars) contains stars in regions of higher or lower density, because one is then not comparing a subset of objects to an `average', such as the low-mass stars (0.08 -- 1.0\,M$_\odot$) in the sample.

We measure $\Sigma_{\rm LDR} = 1.42$ for the objects within 14' of the centre, and as for the full sample, the difference between the surface densities of the most massive stars compared to the whole region are not significant. We find similar conclusions when comparing the surface densities of the substellar objects to the whole region.\\

To summarise the results for IC\,348, the star-forming region is centrally concentrated, with no evidence for mass segregation according to $\Lambda_{\rm MSR}$, but a suggestion that the most massive stars reside in areas of higher than average surface density.

\subsection{NGC\,1333}

\begin{figure*}
  \begin{center}
\setlength{\subfigcapskip}{10pt}
\hspace*{-1.5cm}\subfigure[$\Lambda_{\rm MSR}$, high mass]{\label{ngc1333_data-a}\rotatebox{270}{\includegraphics[scale=0.25]{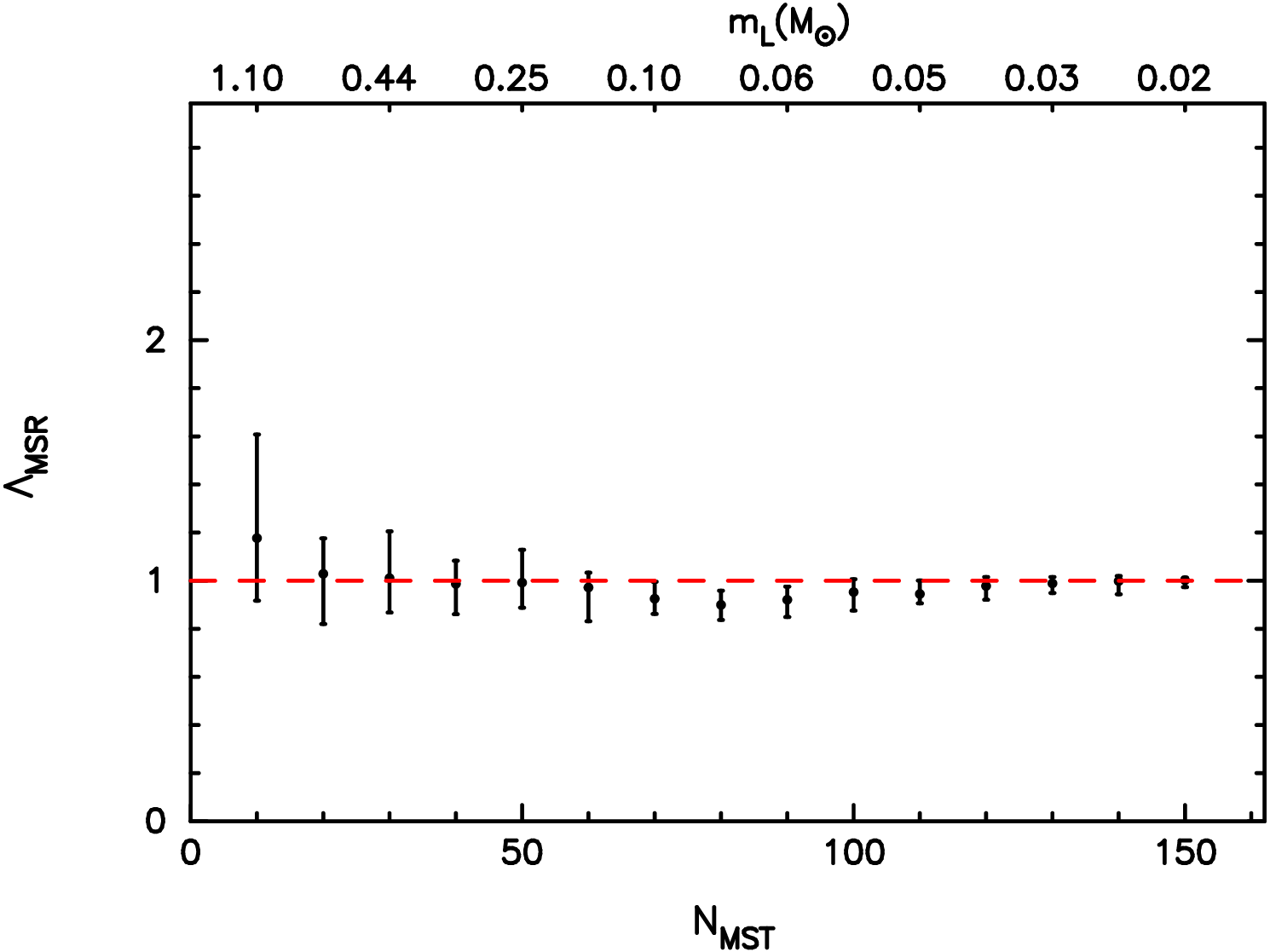}}}
\hspace*{0.3cm} 
\subfigure[$\Lambda_{\rm MSR}$, BDs]{\label{ngc1333_data-b}\rotatebox{270}{\includegraphics[scale=0.24]{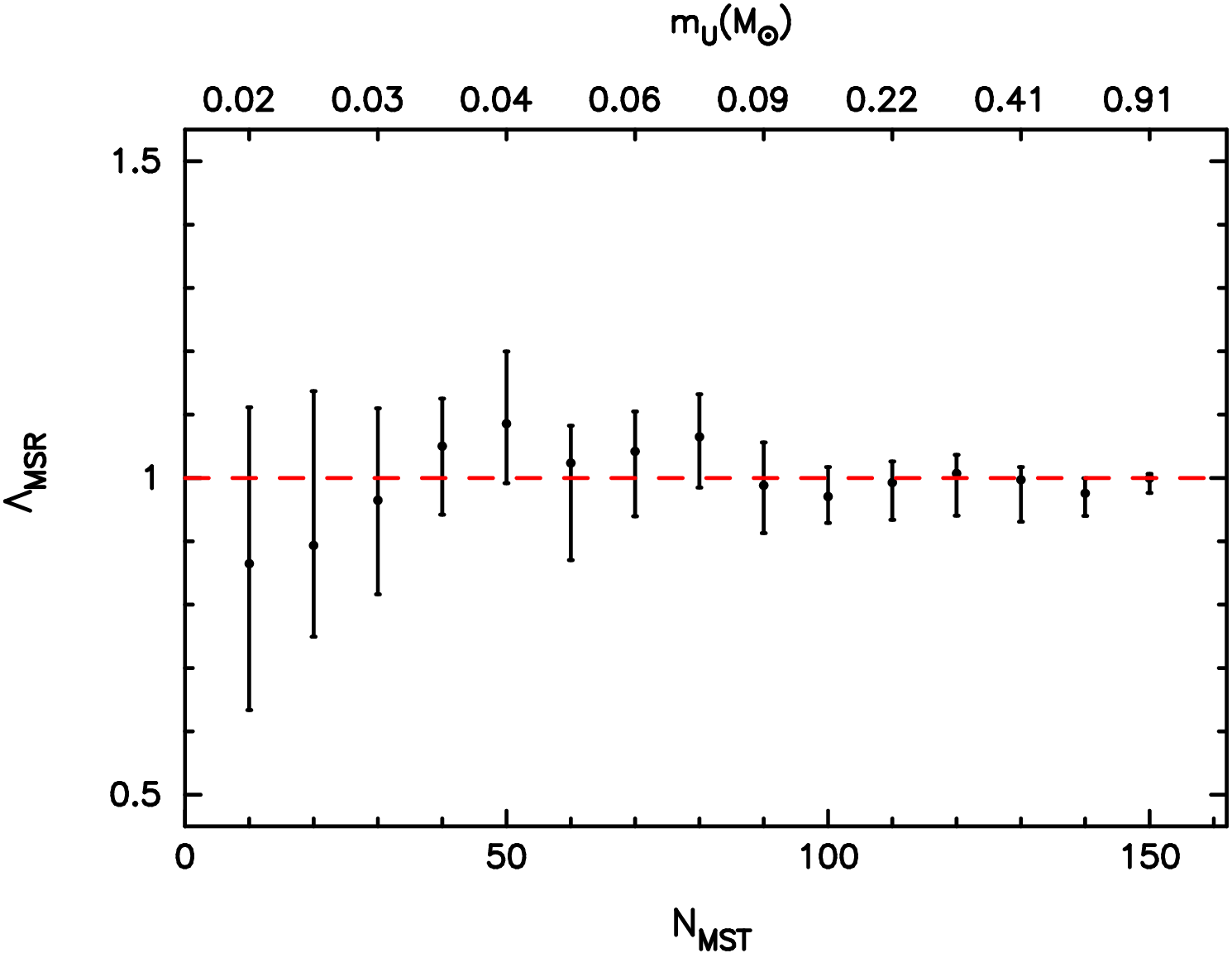}}} 
\hspace*{0.3cm}\subfigure[$\Sigma - m$]{\label{ngc1333_data-c}\rotatebox{270}{\includegraphics[scale=0.25]{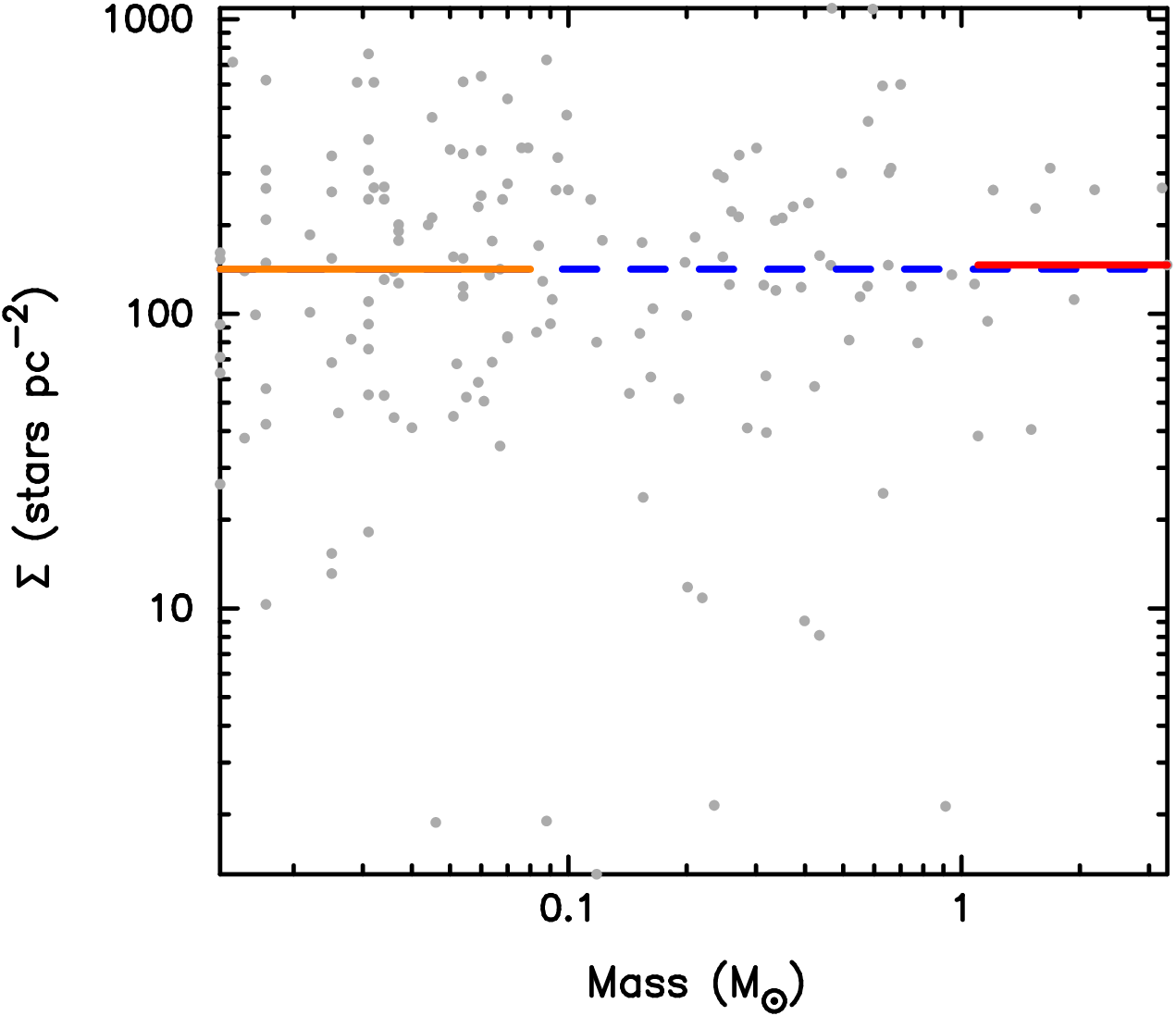}}}
\caption[bf]{Data for NGC1333. Panel (a) shows $\Lambda_{\rm MSR}$ as a function of the $N_{\rm MST}$ most massive stars. $\Lambda_{\rm MSR} = 1.2$, which is probably not a significant deviation from unity. Panel (b) shows $\Lambda_{\rm MSR}$ as a function of the  $N_{\rm MST}$ least massive objects. Again, the deviation from unity is not likely to be significant. Panel (c) shows the median surface density for the most massive stars (red line), cluster average (blue line) least massive BDs (orange line). Deviations from the median are not significant (KS test p-values of 0.92 and 0.22, respectively). The $\mathcal{Q}$-parameter is 0.89 (centrally concentrated).}
\label{ngc1333_data}
  \end{center}
\end{figure*}

\subsubsection{Spatial structure}  

We obtain $\mathcal{Q} = 0.89$, which also indicates a centrally concentrated distribution and $\bar{m} = 0.33$ and $\bar{s} = 0.37$. Due to the lower number of objects (162 compared to 459 for IC\,348), placing constraints on the structure of NGC\,1333 is more difficult, but like IC\,348 it is consistent with a $N \propto r^{-2.5}$ profile. 

If we restrict our determination of the $\mathcal{Q}$-parameter to within the ACIS-I  completeness field, we find $\mathcal{Q} = 0.79$, with $\bar{m} = 0.50$ and $\bar{s} = 0.64$. This is a less centrally concentrated spatial profile than is measured for the full dataset, and corresponds to a uniform distribution. Again, this is unsurprising as we are pruning the more distant objects from our determination of $\mathcal{Q}$.

\subsubsection{Mass segregation ratio, $\Lambda_{\rm MSR}$}

In Fig.~\ref{ngc1333_data-a} we show the evolution of $\Lambda_{\rm MSR}$ as a function of the $N_{\rm MST}$ most massive stars in the subset, which is compared to the MSTs of objects randomly drawn from the spatial distribution shown in Fig.~\ref{ngc1333_map}. $\Lambda_{\rm MSR} = 1.2^{+0.4}_{-0.3}$ for the $N_{\rm MST} = 10$ most massive stars, and also does not deviate from unity if we consider fewer (e.g.\,\,$N_{\rm MST} = 4$) or more stars. 

We then show the $\Lambda_{\rm MSR}$ ratio for the $N_{\rm MST}$ least massive objects (brown dwarfs) in Fig.~\ref{ngc1333_data-b}.  $\Lambda_{\rm MSR} = 0.9^{+0.2}_{-0.3}$ for the $N_{\rm MST} = 10$ least massive brown dwarfs and this plot suggests that these objects are not more or less spatially concentrated than the stars in IC\,348. 

$\Lambda_{\rm MSR} \sim 1$ for the ACIS-I completeness field, suggesting that our results are not influenced by the inclusion or otherwise of outlying objects.

\subsubsection{Local surface density ratio, $\Sigma_{\rm LDR}$}

The local surface density as a function of mass is shown for all of the stars in our sample in Fig.~\ref{ngc1333_data-c}. The median surface density for the whole cluster is shown by the blue dashed line, the median for the ten most massive stars is shown by the red line and the median surface density of the brown dwarfs is shown by the orange line.  The surface density ratio for the most massive stars is $\Sigma_{\rm LDR} = 1.03$, and a KS test between the two populations gives $D = 0.17$ and a $p-$value of 0.92 that the two populations share the same underlying parent distribution and we conclude that the most massive stars do not reside in more dense locations. 

If we compare the surface densities of brown dwarfs to the surface density of the whole star-forming region we have $\Sigma_{\rm LDR} = 1.00$, and the KS test between the two distributions gives $D = 0.33$ and the $p-$value is 0.22, again suggesting that the two populations share the same underlying distribution.

$\Sigma_{\rm LDR} \sim 1$ for both the most massive stars and the substellar objects compared to all objects within the ACIS-I completeness field, suggesting that our results are not influenced by the inclusion or otherwise of outlying objects. \\

To summarise the results for NGC\,1333, the star-forming region is centrally concentrated, with no evidence for mass segregation according to $\Lambda_{\rm MSR}$, and the most massive stars reside in areas of similar surface density to the average.

\section{Comparison with $N$-body simulations}
\label{nbody_comp}

In order to place our values for $\mathcal{Q}$, $\Lambda_{\rm MSR}$ and $\Sigma_{\rm LDR}$ in context, we compare them to snapshots of $N$-body simulations of the evolution of star-forming regions. \citet{Parker14b} and \citet{Parker14e} show that using combinations of  $\mathcal{Q}$, $\Lambda_{\rm MSR}$ and $\Sigma_{\rm LDR}$ can place constraints on the initial virial ratio and the initial density of a star-forming region. 

Both IC\,348 and NGC\,1333 have $\mathcal{Q} > 0.8$, indicative of a smooth, centrally concentrated distribution with no substructure \citep{Cartwright04,Bastian09,Sanchez09}. Because substructure is the expected outcome of star formation in filaments, we would expect dynamically pristine regions (i.e.\,\,regions that have experienced little or no dynamical evolution) to have  $\mathcal{Q} << 0.8$. This is corroborated by hydrodynamical simulations of star formation, which almost always display significant substructure \citep{Bate09,Girichidis12,Dale14,Parker15a}. It is possible to `freeze in' substructure if the initial virial state of the region is unbound, or if the region is in virial equilibrium  but has a low stellar density \citep{Parker14b}. 

Given that neither IC\,348 nor NGC\,1333 display substructure, we can deduce that they were initially subvirial or virialised, and so we will compare the observed datapoints to simulations of star-forming regions undergoing sub-virial collapse to form a cluster \citep{Adams06,Allison10,Parker14b}.

\subsection{Simulation initial conditions}

We tailor our $N$-body simulations to the observed star-forming regions by assigning masses from the complete samples for each region from \citet{Luhman16}, i.e.\,\,the objects within the 14' radius in IC\,348 and the objects within the ACIS-I field in NGC\,1333. This gives a total of 425 objects for our IC\,348-like regions, and 150 objects for NGC\,1333. We randomly assign masses to the objects in each simulation, meaning there is no preferred spatial distribution initially for any mass range. The mass functions for both regions are consistent with the \citet{Maschberger13} initial mass function (IMF), and in Appendix~\ref{appendix} we show how changing the mass function has little effect on the dynamical evolution of the regions.

The simulations are set up with primordial substructure, in an effort to mimic the initial conditions of star formation in both observed and (hydrodynamical) simulated star-forming regions. We create substructured regions using the fractal method, which is a (relatively) straightforward way of setting up substructure and we refer the interested reader to \citet{Goodwin04a,Parker14b} for full details of the method. Observations \citep[e.g.][]{Larson81,Hacar13,Andre14} suggest that stellar velocities are correlated on local scales at formation, and our fractal generator creates kinematic substructure as well as spatial substructure \citep[see][for details]{Parker16b}. The velocities of the stars are then scaled to the global virial ratio, $\alpha_{\rm vir} = 0.3$, which makes the regions subvirial and induces collapse \citep{Adams06,Allison10}.

The degree of spatial and kinematic substructure is set by the fractal dimension, $D$, and our simulations all have $D = 1.6$ initially \citep[see][for a full investigation of substructure parameter space]{Parker14b}. It is not straightforward to directly measure the fractal dimension of a star-forming region, and a proxy is usally used. We use the $\mathcal{Q}$-parameter \citep{Cartwright04,Cartwright09b,Lomax11,Jaffa17}, which can be used to infer the fractal dimension of a region. However, the $\mathcal{Q}$-parameter is slightly dependent on the number of objects in a sample \citep{Lomax11,Parker15c}, even though it is independent of the global size and density of a region. As an example, a highly substructured region ($D = 1.6$) with 425 stars will have a lower  $\mathcal{Q}$-parameter than a region with 150 stars, and this should be kept in mind when comparing star-forming regions with different numbers of objects \citep[see e.g.][their fig.~1 for an illustration of this]{Parker15c}.

We set up our simulations with a range of radii so as to vary the initial \emph{local} stellar density. Depending on the simulation, this results in regions with initial densities as low as 10\,M$_\odot$\,pc$^{-3}$ to densities as high as 5000\,M$_\odot$\,pc$^{-3}$. A summary of the simulations, including their initial radii, stellar density and number of stars is given in Table~\ref{simulations}.

\begin{table*}
  \caption[bf]{Summary of the $N$-body simulations used in this work. The columns show the star-forming region, mumber of stars, the IMF used (either the data from \citet{Luhman16} or the \citet{Maschberger13} IMF), the radius of the star-forming region, $r_F$ and the median initial local density this radius results in, $\tilde{\rho}_{\rm ini}$.}
  \begin{center}
    \begin{tabular}{|c|c|c|c|c|}
      \hline
      Region & $N_{\rm stars}$ & IMF & $r_F$ & $\tilde{\rho}_{\rm ini}$  \\
      \hline
      IC\,348 & 425 & Observed \citep{Luhman16} & 0.5\,pc & 5000 -- 10\,000\,M$_\odot$\,pc$^{-3}$ \\
      IC\,348 & 425 & Observed \citep{Luhman16} & 1.5\,pc & 100 -- 500\,M$_\odot$\,pc$^{-3}$ \\
      IC\,348 & 425 & \citet{Maschberger13} & 1.5\,pc & 100 -- 500\,M$_\odot$\,pc$^{-3}$ \\
      IC\,348 & 425 & Observed \citep{Luhman16} & 3\,pc & 10 -- 60\,M$_\odot$\,pc$^{-3}$ \\
      \hline
      NGC\,1333 & 150 & Observed \citep{Luhman16} & 0.5\,pc & 500 -- 2000\,M$_\odot$\,pc$^{-3}$ \\
      NGC\,1333 & 150 & \citet{Maschberger13} & 0.5\,pc & 500 -- 2000\,M$_\odot$\,pc$^{-3}$ \\
      \hline
    \end{tabular}
  \end{center}
  \label{simulations}
\end{table*}

\begin{figure*}
  \begin{center}
    \setlength{\subfigcapskip}{10pt}
    \hspace*{-1.3cm}\subfigure[$\mathcal{Q}$-parameter]{\label{ic348_nbody_high-a}\rotatebox{270}{\includegraphics[scale=0.29]{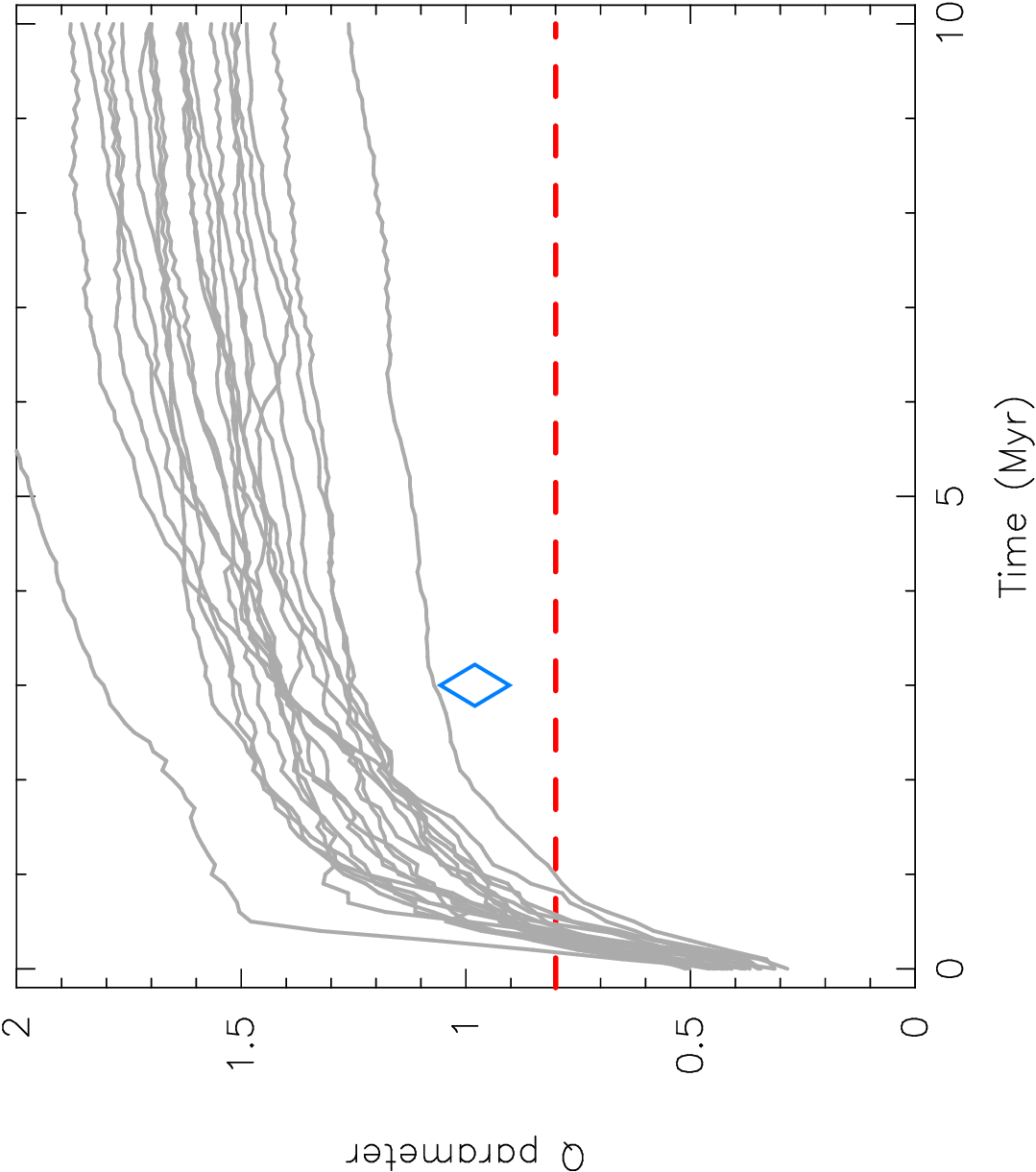}}}
    \hspace*{0.3cm}
    \subfigure[$\mathcal{Q}-\Lambda_{\rm MSR}$]{\label{ic348_nbody_high-b}\rotatebox{270}{\includegraphics[scale=0.29]{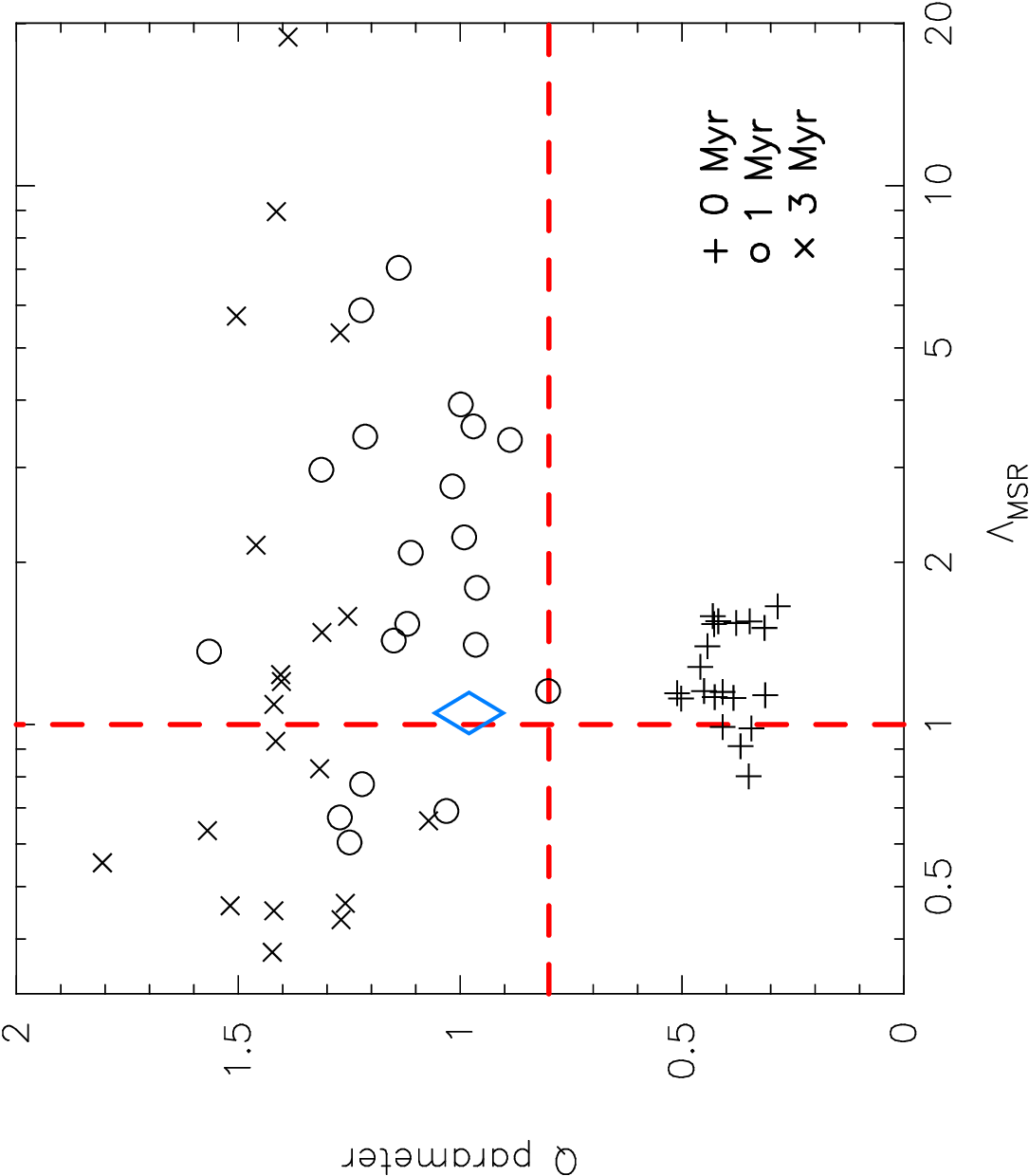}}}
    \hspace*{0.3cm}\subfigure[$\mathcal{Q}-\Sigma_{\rm LDR}$]{\label{ic348_nbody_high-c}\rotatebox{270}{\includegraphics[scale=0.29]{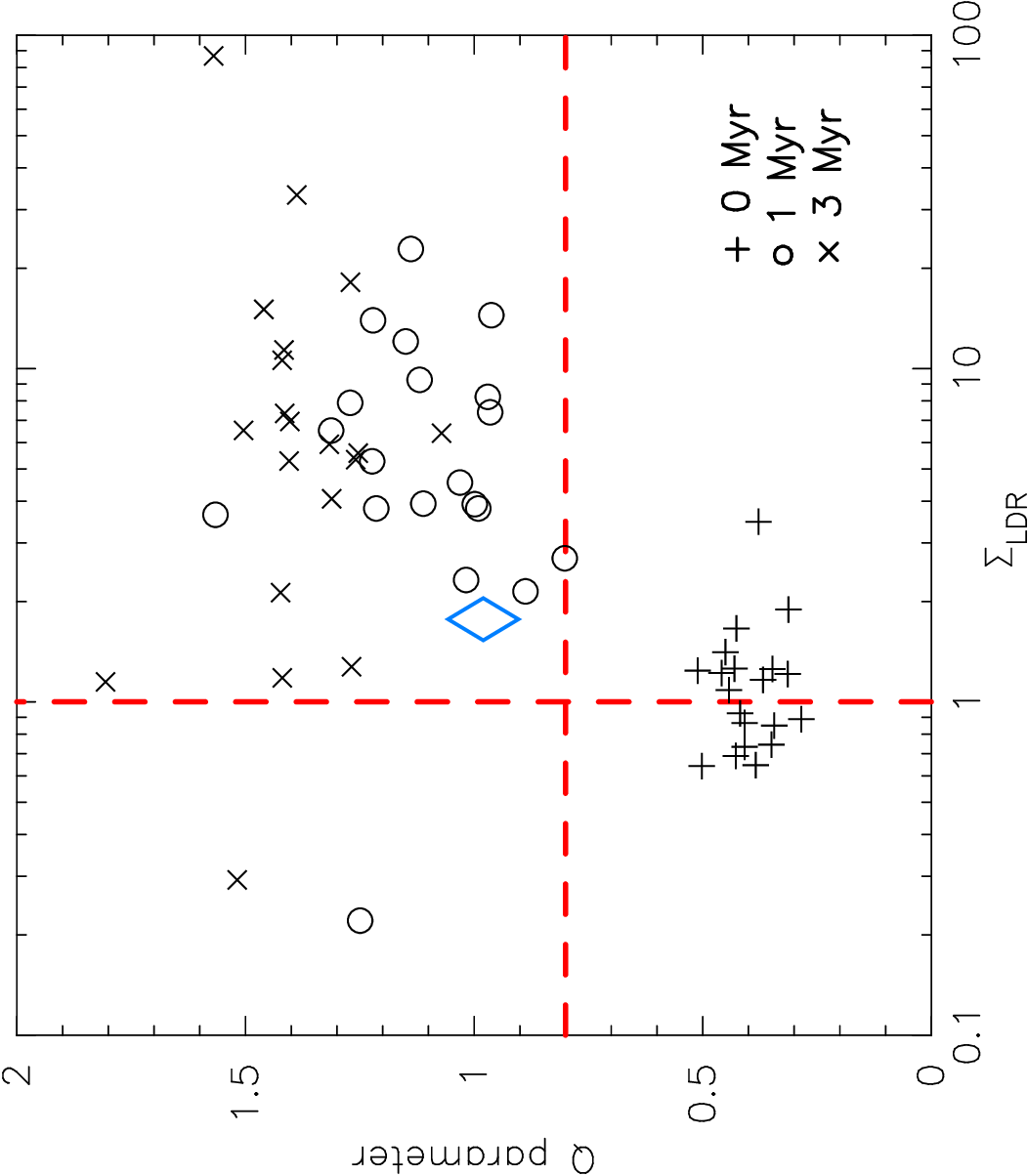}}}
    \caption[bf]{$N$-body simulations of star-forming regions containing 425 stars undergoing cool-collapse with stellar masses from the \citet{Luhman16} sample for IC\,348 and initial stellar densities of 5000 -- 10\,000\,M$_\odot$\,pc$^{-3}$. In panel (a) we show the evolution of the $\mathcal{Q}$-parameter; in  panel (b) we show the $\mathcal{Q}-\Lambda_{\rm MSR}$ plot and in panel (c) we show the $\mathcal{Q}-\Sigma_{\rm LDR}$ plot. in panels (b) and (c) the simulation data at 0, 1 and 3\,Myr are indicated by the plus signs, open circles and crosses, respectively. In all three panels, the blue diamond symbol shows the observational value at 3\,Myr. The dashed horizontal line in all panels indicates the boundary between a substructured ($\mathcal{Q} < 0.8$) and a smooth ($\mathcal{Q} > 0.8$) distribution, and in panels (b) and (c) the vertical dashed lines indicate $\Lambda_{\rm MSR} = 1$ (no mass segregation) and $\Sigma_{\rm LDR} = 1$ (no enhanced surface densities) respectively.}
    \label{ic348_nbody_high}
  \end{center}
\end{figure*}

\begin{figure*}
  \begin{center}
    \setlength{\subfigcapskip}{10pt}
    \hspace*{-1.3cm}\subfigure[$\mathcal{Q}$-parameter]{\label{ic348_nbody_med-a}\rotatebox{270}{\includegraphics[scale=0.29]{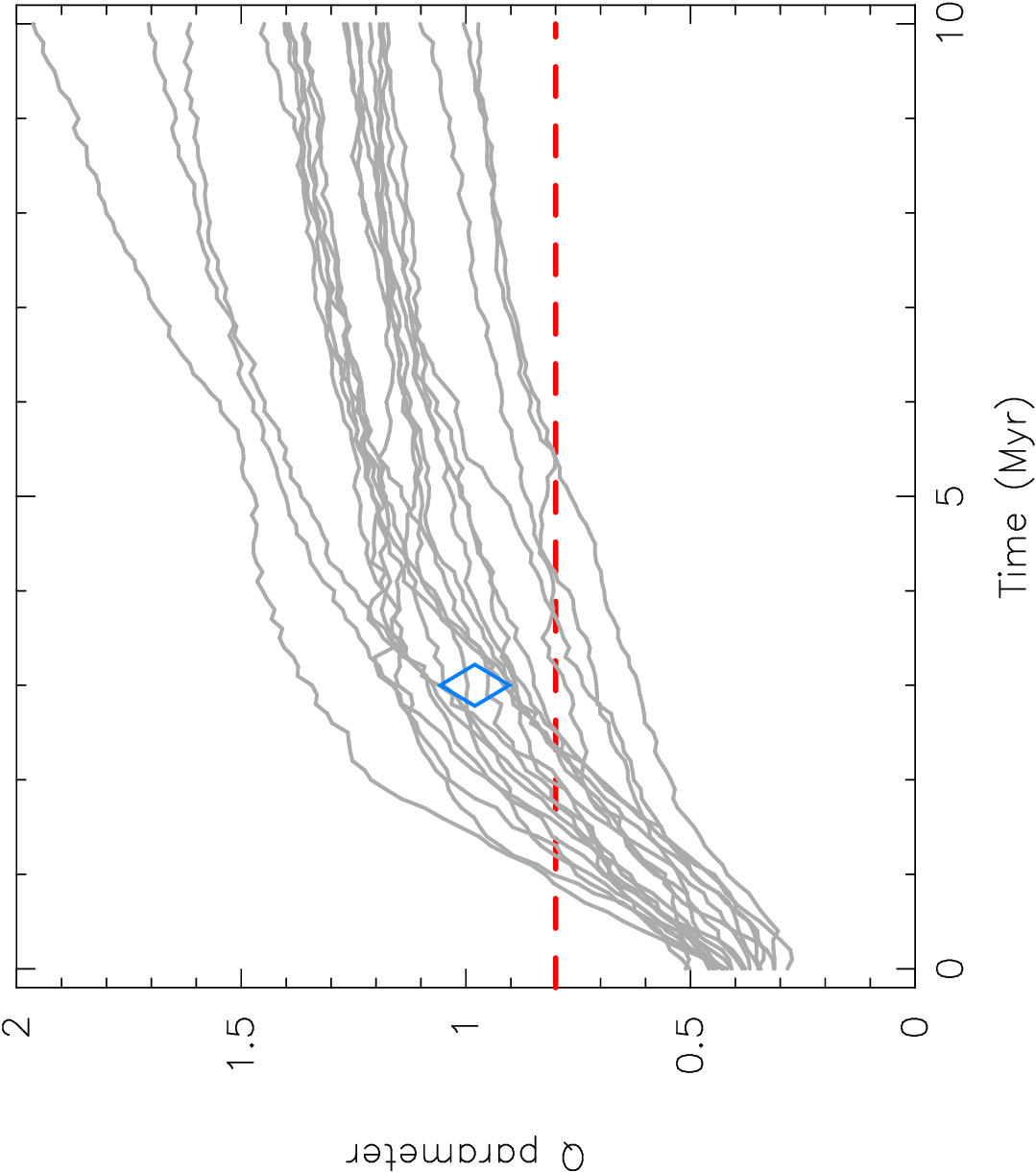}}}
    \hspace*{0.3cm}
    \subfigure[$\mathcal{Q}-\Lambda_{\rm MSR}$]{\label{ic348_nbody_med-b}\rotatebox{270}{\includegraphics[scale=0.29]{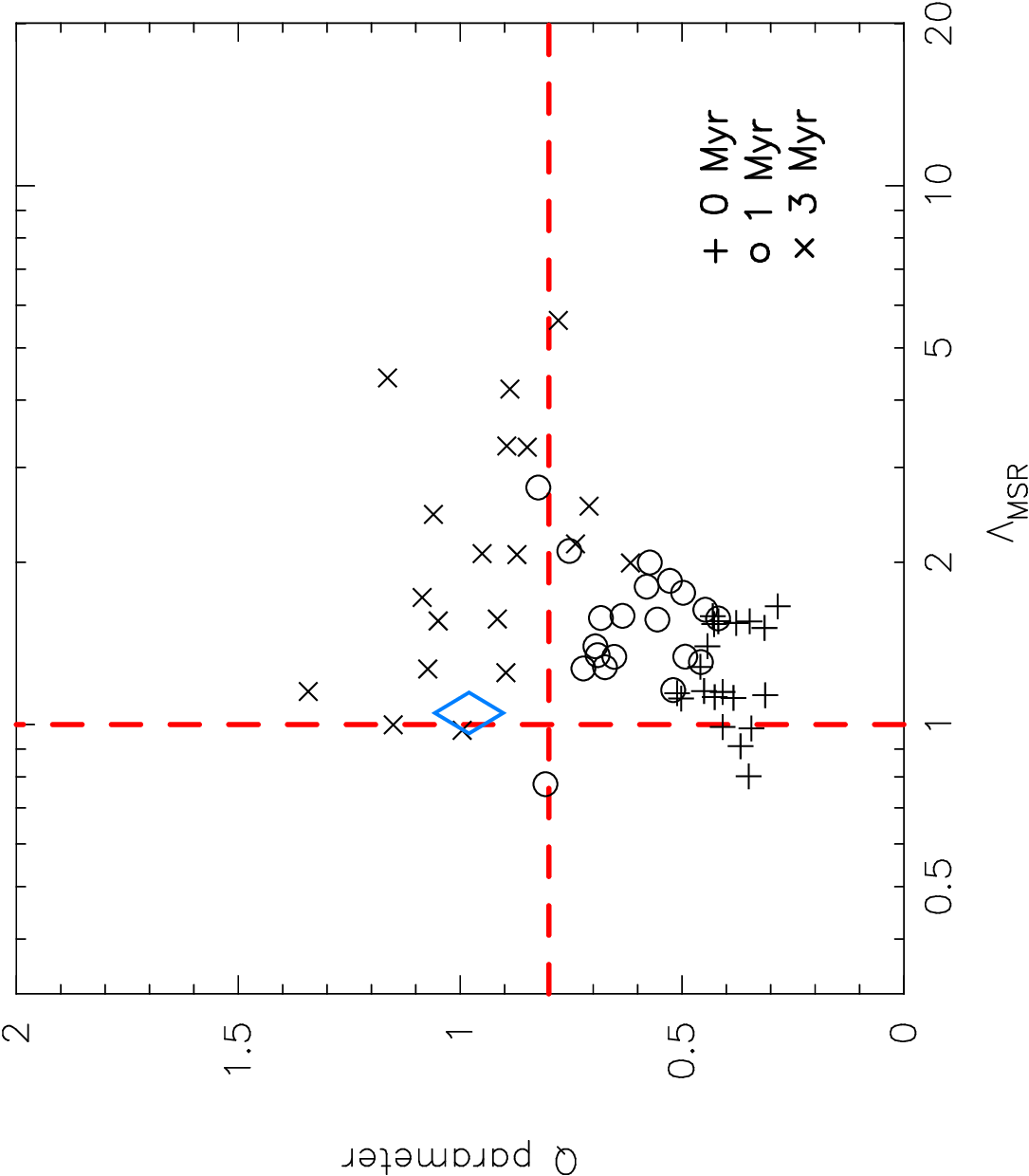}}}
    \hspace*{0.3cm}\subfigure[$\mathcal{Q}-\Sigma_{\rm LDR}$]{\label{ic348_nbody_med-c}\rotatebox{270}{\includegraphics[scale=0.29]{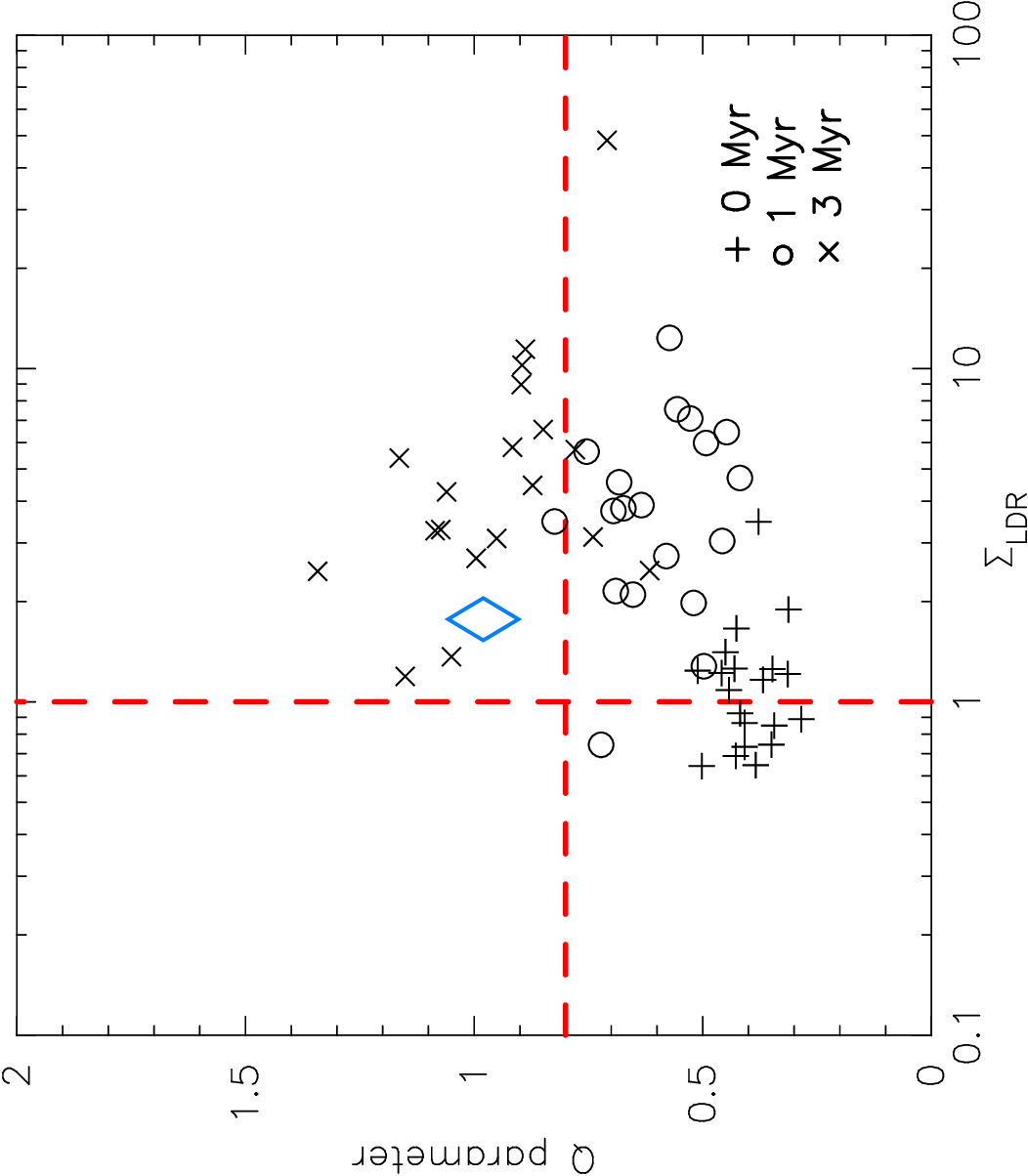}}}
    \caption[bf]{$N$-body simulations of star-forming regions containing 425 stars undergoing cool-collapse with stellar masses from the \citet{Luhman16} sample for IC\,348 and initial stellar densities of 100 -- 500\,M$_\odot$\,pc$^{-3}$. In panel (a) we show the evolution of the $\mathcal{Q}$-parameter; in panel (b) we show the $\mathcal{Q}-\Lambda_{\rm MSR}$ plot and in panel (c) we show the $\mathcal{Q}-\Sigma_{\rm LDR}$ plot. in panels (b) and (c) the simulation data at 0, 1 and 3\,Myr are indicated by the plus signs, open circles and crosses, respectively. In all three panels, the blue diamond symbol shows the observational values at 3\,Myr. The dashed horizontal line in all panels indicates the boundary between a substructured ($\mathcal{Q} < 0.8$) and a smooth ($\mathcal{Q} > 0.8$) distribution, and in panels (b) and (c) the vertical dashed lines indicate $\Lambda_{\rm MSR} = 1$ (no mass segregation) and $\Sigma_{\rm LDR} = 1$ (no enhanced surface densities) respectively.}
        \label{ic348_nbody_med}
  \end{center}
\end{figure*}

\begin{figure*}
  \begin{center}
    \setlength{\subfigcapskip}{10pt}
    \hspace*{-1.3cm}\subfigure[$\mathcal{Q}$-parameter]{\label{ic348_nbody_low-a}\rotatebox{270}{\includegraphics[scale=0.29]{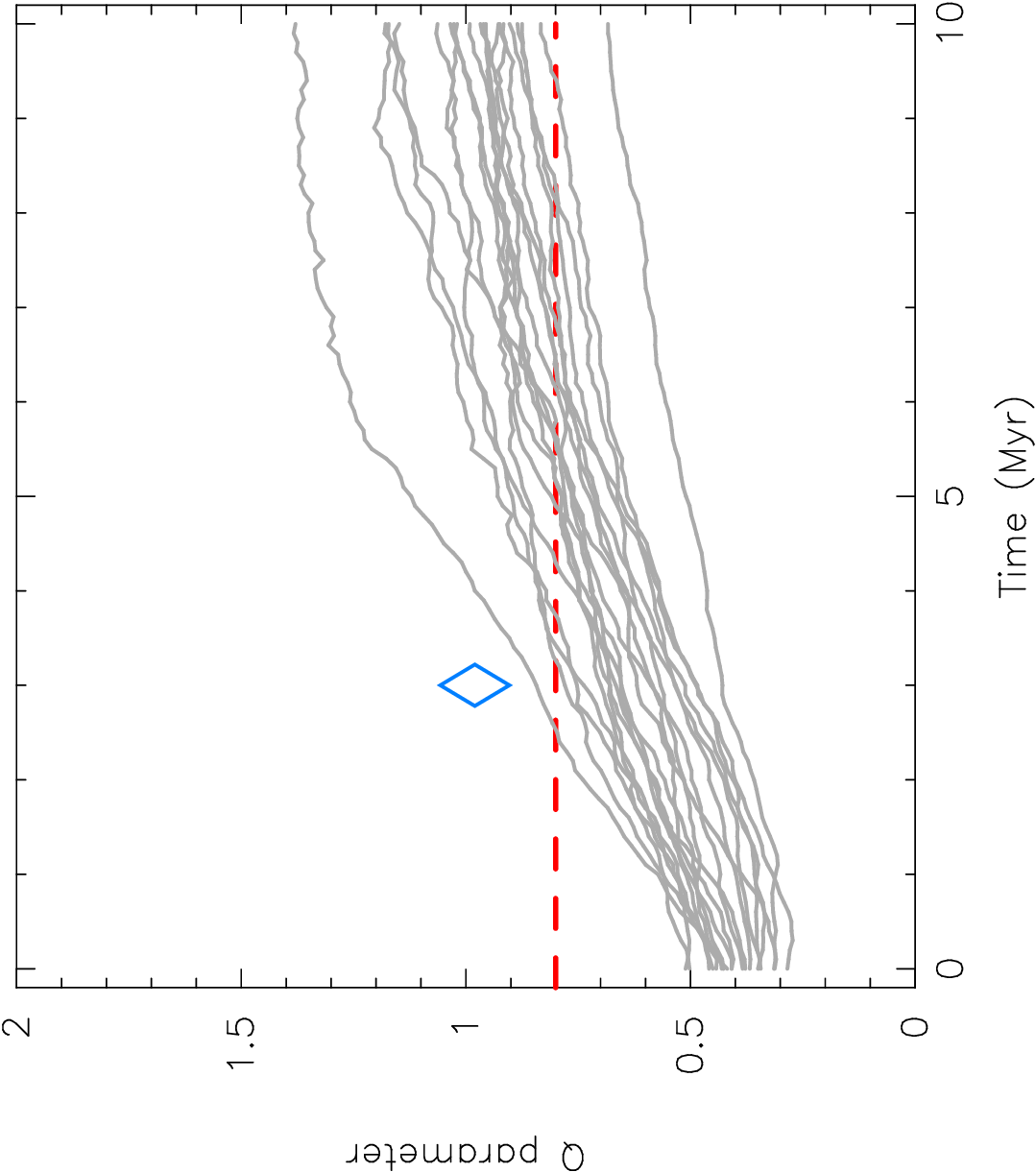}}}
    \hspace*{0.3cm}
    \subfigure[$\mathcal{Q}-\Lambda_{\rm MSR}$]{\label{ic348_nbody_low-b}\rotatebox{270}{\includegraphics[scale=0.29]{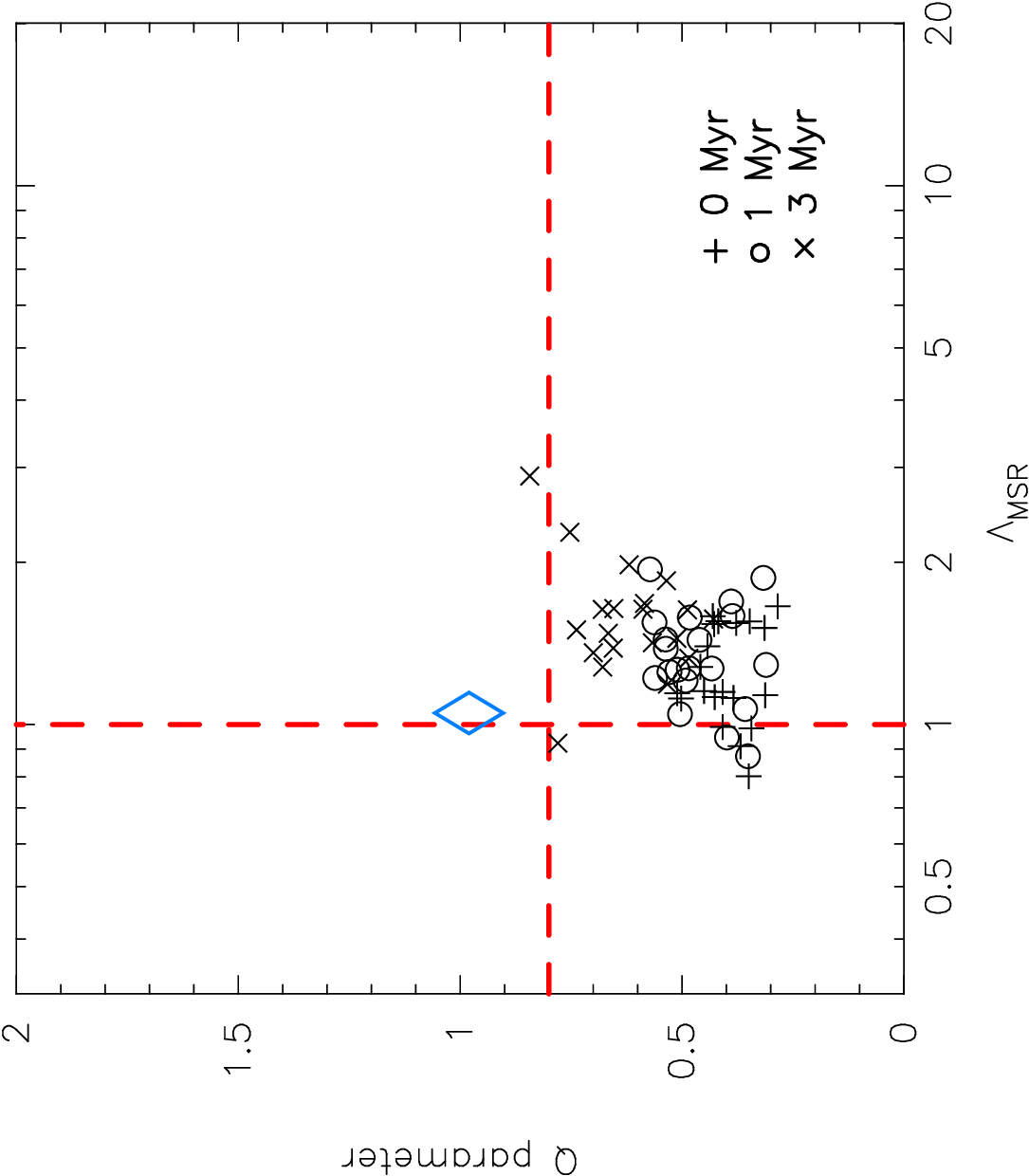}}}
    \hspace*{0.3cm}\subfigure[$\mathcal{Q}-\Sigma_{\rm LDR}$]{\label{ic348_nbody_low-c}\rotatebox{270}{\includegraphics[scale=0.29]{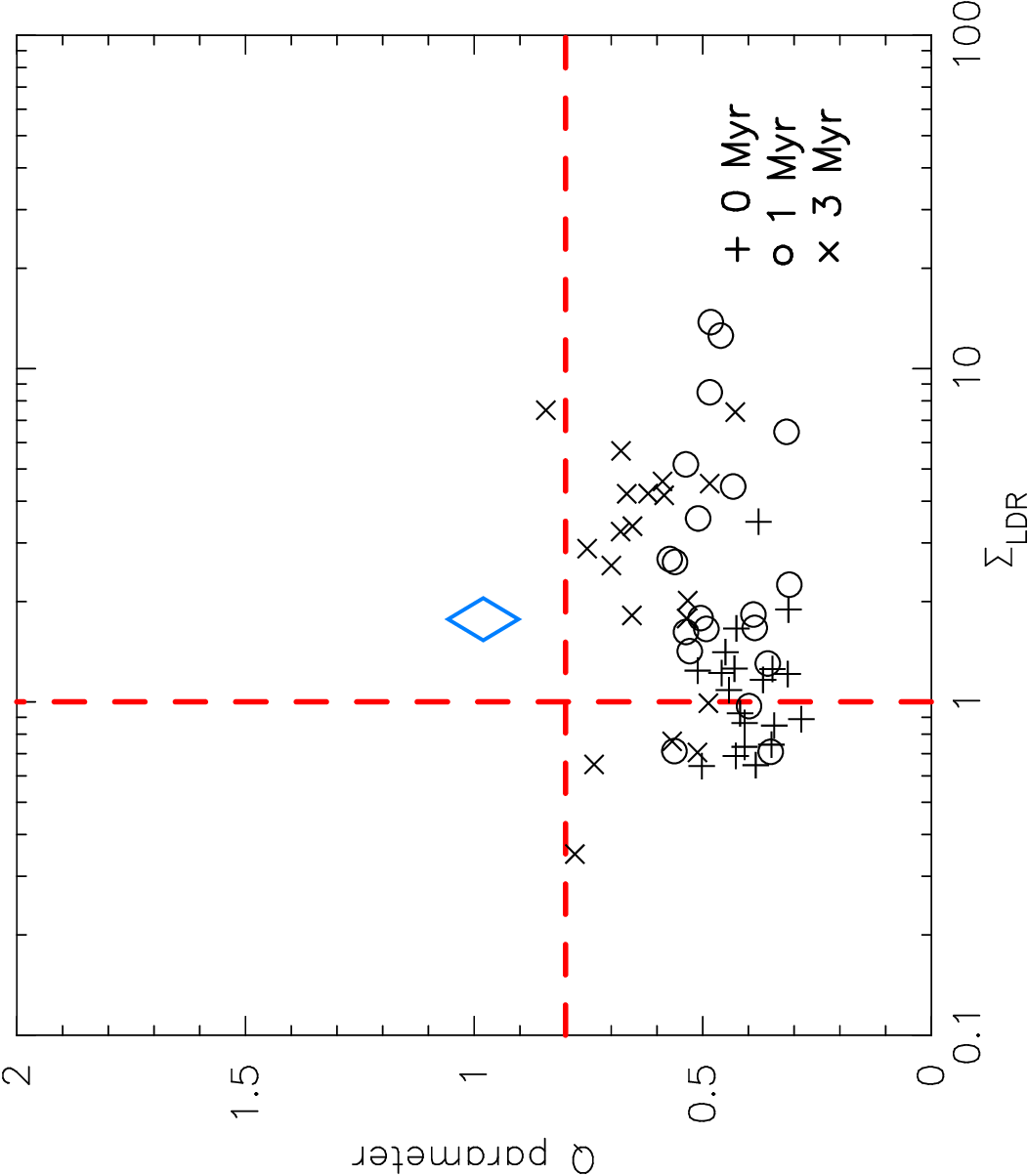}}}
    \caption[bf]{$N$-body simulations of star-forming regions containing 425 stars undergoing cool-collapse with stellar masses from the \citet{Luhman16} sample for IC\,348 and initial stellar densities of 10 -- 60\,M$_\odot$\,pc$^{-3}$. In panel (a) we show the evolution of the $\mathcal{Q}$-parameter; in panel (b) we show the $\mathcal{Q}-\Lambda_{\rm MSR}$ plot and in panel (c) we show the $\mathcal{Q}-\Sigma_{\rm LDR}$ plot. in panels (b) and (c) the simulation data at 0, 1 and 3\,Myr are indicated by the plus signs, open circles and crosses, respectively. In all three panels, the blue diamond symbol shows the observational value at 3\,Myr. The dashed horizontal line in all panels indicates the boundary between a substructured ($\mathcal{Q} < 0.8$) and a smooth ($\mathcal{Q} > 0.8$) distribution, and in panels (b) and (c) the vertical dashed lines indicate $\Lambda_{\rm MSR} = 1$ (no mass segregation) and $\Sigma_{\rm LDR} = 1$ (no enhanced surface densities) respectively.}
    \label{ic348_nbody_low}
  \end{center}
\end{figure*}

\begin{figure*}
  \begin{center}
    \setlength{\subfigcapskip}{10pt}
    \hspace*{-1.3cm}\subfigure[$\mathcal{Q}$-parameter]{\label{ngc1333_nbody-a}\rotatebox{270}{\includegraphics[scale=0.29]{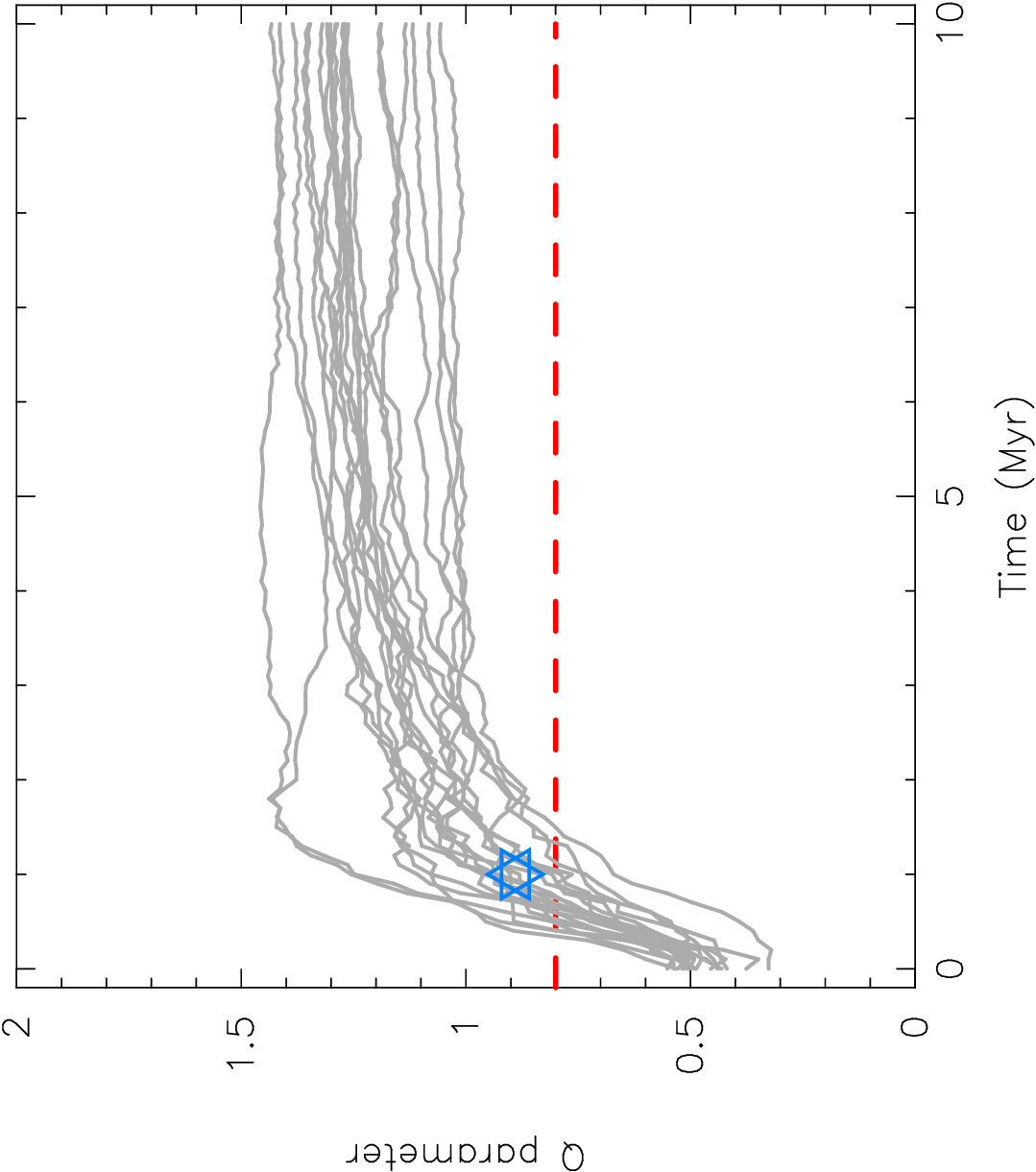}}}
    \hspace*{0.3cm}
    \subfigure[$\mathcal{Q}-\Lambda_{\rm MSR}$]{\label{ngc1333_nbody-b}\rotatebox{270}{\includegraphics[scale=0.29]{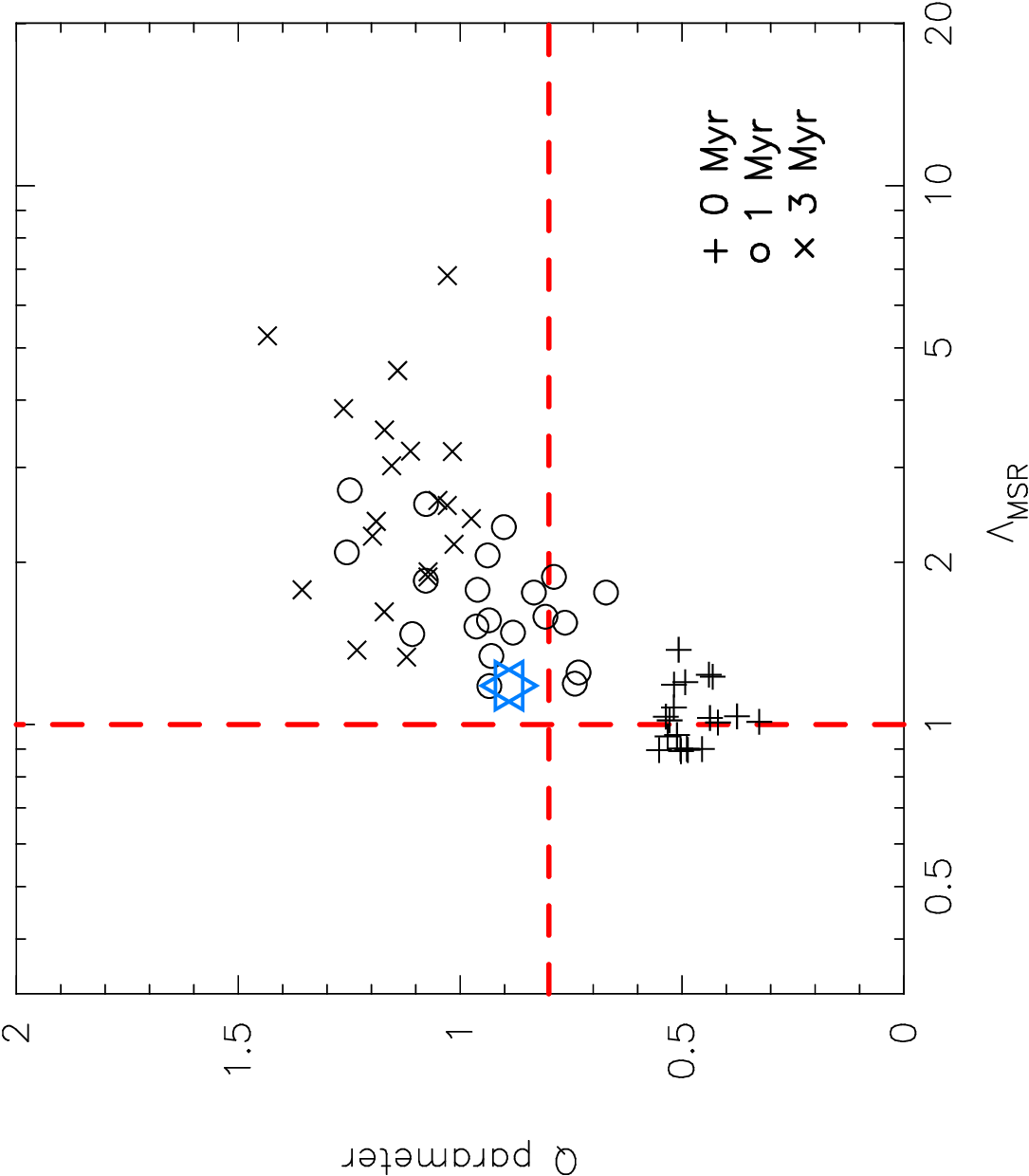}}}
    \hspace*{0.3cm}\subfigure[$\mathcal{Q}-\Sigma_{\rm LDR}$]{\label{ngc1333_nbody-c}\rotatebox{270}{\includegraphics[scale=0.29]{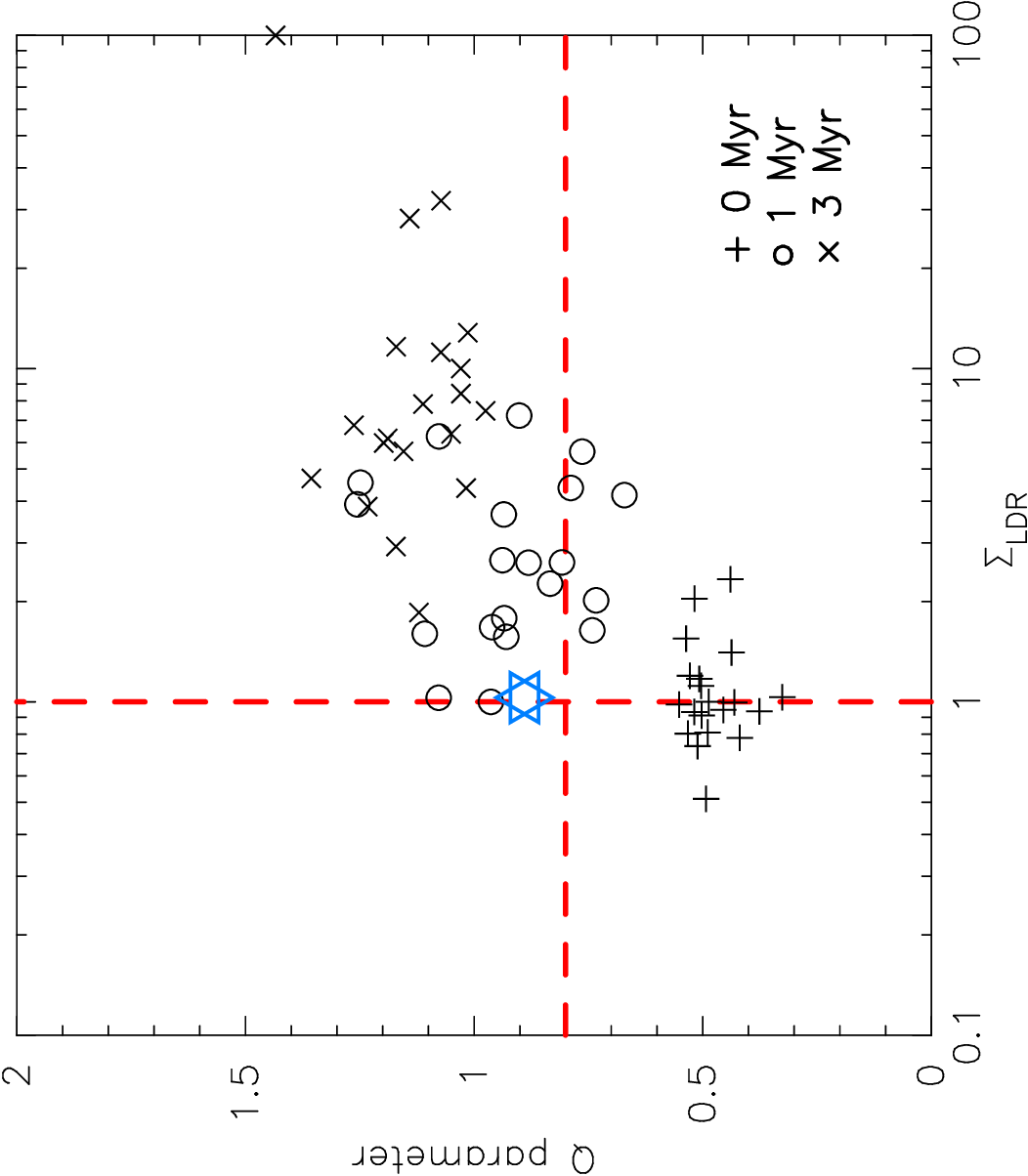}}}
    \caption[bf]{$N$-body simulations of star-forming regions containing 150 stars undergoing cool-collapse with stellar masses from the \citet{Luhman16} sample for NGC\,1333 and initial stellar densities of 500 -- 2000\,M$_\odot$\,pc$^{-3}$. In panel (a) we show the evolution of the $\mathcal{Q}$-parameter; in panel (b) we show the $\mathcal{Q}-\Lambda_{\rm MSR}$ plot and in panel (c) we show the $\mathcal{Q}-\Sigma_{\rm LDR}$ plot. in panels (b) and (c) the simulation data at 0, 1 and 3\,Myr are indicated by the plus signs, open circles and crosses, respectively. In all three panels, the blue hexagram symbol shows the observational value at 1\,Myr. The dashed horizontal line in all panels indicates the boundary between a substructured ($\mathcal{Q} < 0.8$) and a smooth ($\mathcal{Q} > 0.8$) distribution, and in panels (b) and (c) the vertical dashed lines indicate $\Lambda_{\rm MSR} = 1$ (no mass segregation) and $\Sigma_{\rm LDR} = 1$ (no enhanced surface densities) respectively.}
    \label{ngc1333_nbody}
  \end{center}
\end{figure*}

The simulations are evolved for 10\,Myr using the \texttt{kira} integrator in the \texttt{Starlab} environment \citep{Zwart99,Zwart01}. We do not include stellar evolution in the simulations as neither IC\,348 or NGC\,1333 contain stars massive enough to evolve significantly in the first 10\,Myr. 

\subsection{Comparison with IC\,348}

In Figs.~\ref{ic348_nbody_high},~\ref{ic348_nbody_med}~and~\ref{ic348_nbody_low} we show the simulations set up to constrain the initial density of IC\,348. In all three figures the blue diamond symbol corresponds to the observed value for IC\,348 at its current age (3\,Myr). In each figure, panel (a) shows the evolution of the $\mathcal{Q}$-parameter, panel (b) shows the $\mathcal{Q}$-parameter against the mass segregation ratio $\Lambda_{\rm MSR}$ at three epochs in the simulation (0, 1 and 3\,Myr) and panel (c) shows the $\mathcal{Q}$-parameter against the local surface density ratio $\Sigma_{\rm LDR}$ at the same three epochs. The boundary between a substructured and a smooth distribution is indicated by the horizontal dashed line in all panels, and in panels (b) and (c) $\Lambda_{\rm MSR} = 1$ (no mass segregation) and $\Sigma_{\rm LDR} = 1$ (no enhanced surface densities) are shown by the vertical dashed lines.                                                                 

We show the evolution of simulations with initial radii $r_F = 0.5$\,pc and 425 objects with masses from the \citet{Luhman16} dataset assigned randomly with position in Fig.~\ref{ic348_nbody_high}. Our fractal initial conditions with this radius gives an initial median stellar density between 5000 -- 10\,000\,M$_\odot$\,pc$^{-3}$, depending on the random number seed used to initialise the positions of the objects. These high stellar densities facilitate dynamical interactions, and the $\mathcal{Q}$-parameters in the simulations are higher at 3\,Myr than the observed value because interactions erase substructure \citep{Scally02,Goodwin04a,Parker12d}. Similarly, due to the high number of interactions, dynamical mass segregation often occurs (panel b), which is a natural outcome of dense, substructured and (sub)virial initial conditions \citep{McMillan07,Allison10,Parker14b}.

The massive stars also attain higher than average local densities (panel c) because they typically migrate to the centre of a subvirial, collapsing star-forming region \citep{Parker14b}, although other initial conditions can lead to high $\Sigma_{\rm LDR}$ values and for example the $\mathcal{Q} - \Sigma_{\rm LDR}$ plot (panel c) can used to distinguish between a dense, subvirial region (which would have high $\Sigma_{\rm LDR}$ and high $\Lambda_{\rm MSR}$) and a dense, supervirial ($\alpha_{\rm vir} = 1.5$), whch would have high $\Sigma_{\rm LDR}$ but $\Lambda_{\rm MSR} \sim 1$. The evolution of a cluster with dense, subvirial initial conditions is to move up and to the right in both the $\mathcal{Q} - \Lambda_{\rm MSR}$ and $\mathcal{Q} - \Sigma_{\rm LDR}$ plots as it erases substructure, dynamically mass segregates and the massive stars attain higher than average local densities, all of which are inconsistent with the observed values (the blue diamonds).

In Fig.~\ref{ic348_nbody_med} we show the evolution of simulations with initial radii $r_F = 1.5$\,pc, 425 objects and masses assigned from the \citet{Luhman16} sample, which gives initial stellar densities between 100 -- 500\,M$_\odot$\,pc$^{-3}$. These stellar densities lead to a moderate level of dynamical evolution; substructure is erased within the first 1 -- 5\,Myr, and the simulations are consistent with the observed $\mathcal{Q}$-parameter at ages of 3\,Myr (panel a). Around half of our simulations display mass segregation after 3\,Myr, but several are consistent with the observed $\Lambda_{\rm MSR}$ value \emph{and} $\mathcal{Q}$-parameter at these ages (the crosses in panel b). Similarly, several simulations present higher $\Sigma_{\rm LDR}$ values than are observed (panel c), but 25\,per cent of the simulations are close to the observed $\mathcal{Q} - \Sigma_{\rm LDR}$ value on the plot.

In Fig.~\ref{ic348_nbody_low} we show the evolution of simulations with initial radii $r_F = 3$\,pc, 425 objects and masses assigned from the \citet{Luhman16} sample, which gives initial stellar densities between 10 -- 60\,M$_\odot$\,pc$^{-3}$. Spatial substructure is maintained due to these low densities, and the regions remain substructured until well after 5\,Myr (panel a). No significant mass segregation occurs (panel b), but some simulations display high local surface densities for the most massive stars ($\Sigma_{\rm LDR} >> 1$) after 3\,Myr. However, because of the retention of substructure, $\mathcal{Q} < 1$ and the $\mathcal{Q} - \Lambda_{\rm MSR}$ and $\mathcal{Q} - \Sigma_{\rm LDR}$ values are inconsistent with the observations. 

In summary, our $N$-body simulations of IC\,348 suggest the region was initially substructured and sub-virial or in virial equilibrium, with an initial stellar density in the range of 100 -- 500\,M$_\odot$\,pc$^{-3}$. 

\subsection{Comparison with NGC\,1333}

We now focus on $N$-body simulations that attempt to mimic NGC\,1333. We ran a series of models with different initial stellar densities, but for the sake of brevity we omit a discussion of the models whose initial density leads to dynamical evolution that is not consistent with the observations.

In Fig.~\ref{ngc1333_nbody} we show the evolution of simulations with initial radii $r_F = 0.5$\,pc, 150 objects and masses assigned from the \citet{Luhman16} sample, which gives initial stellar densities between 500 -- 2000\,M$_\odot$\,pc$^{-3}$. These relatively high stellar densities cause the erasure of substructure within the first 1 -- 2\,Myr, as shown by the evolution of the $\mathcal{Q}$-parameter in Fig.~\ref{ngc1333_nbody-a}. However, the simulations rarely show evidence of mass segregation (panel b) or enhanced surface densities for the most massive stars (panel c) within the first 1\,Myr. We find that simulations with much higher ($>$5000\,M$_\odot$\,pc$^{-3}$) or lower ($<$100\,M$_\odot$\,pc$^{-3}$) densities are inconsistent with the observations. The boundary between a substructured and a smooth distribution is indicated by the horizontal dashed line in all panels, and in panels (b) and (c) $\Lambda_{\rm MSR} = 1$ (no mass segregation) and $\Sigma_{\rm LDR} = 1$ (no enhanced surface densities) are shown by the vertical dashed lines. The blue hexagram symbols indicate the observed values for NGC\,1333 assuming an age of 1\,Myr.

The simulation values at 1\,Myr are therefore consistent with the observed value for NGC\,1333 if the region was initially substructured and sub-virial or in virial equilibrium, with an initial stellar density in the range of 500 -- 2000\,M$_\odot$\,pc$^{-3}$.

\section{Discussion}
\label{discuss}

Our spatial analysis of IC\,348 and NGC\,1333 show that both star-forming regions are centrally concentrated, with no mass segregation and no enhanced surface densities for the most massive stars. The comparison with the $N$-body models suggest that IC\,348 is consistent with having formed with some substructure, subvirial initial conditions and moderate stellar densities ($\tilde{\rho} \sim 100 - 500$\,M$_\odot$\,pc$^{-3}$). NGC\,1333 is likely to have formed with slightly denser initial conditions ($\tilde{\rho} \sim 500 - 2000$\,M$_\odot$\,pc$^{-3}$).

\subsection{Potential biases}

In order to determine the structural diagnostics $\mathcal{Q}$, $\Lambda_{\rm MSR}$ and $\Sigma_{\rm LDR}$ in a region, we must first be confident of  membership for each star (all three diagnostics), and secondly, reasonable estimates for individual stellar masses must be made ($\Lambda_{\rm MSR}$ and $\Sigma_{\rm LDR}$).

The datasets from \citet{Luhman16} are the most complete censuses of NGC\,1333 and IC\,348 to date, and therefore the $\mathcal{Q}$-parameters are unlikely to be affected by incompleteness \citep{Parker12d,Parker15c}. However, as we have shown, when we restrict our analyses to the regions that have been extensively surveyed (within the 14' radius in IC\,348 and the ACIS-I field in NGC\,1333), the $\mathcal{Q}$-parameters are lower due to the pruning of more distant objects from the analysis. It is important to emphasise that if the areas outside the completeness limits are incomplete, and there is therefore more substructure in both regions, then each region is even more dynamically young, which reinforces our conclusion that they formed with moderate stellar densities. There is very little difference in the values of $\Lambda_{\rm MSR}$ and $\Sigma_{\rm LDR}$ within the completeness fields compared to the full datasets, which again suggests that our conclusions are robust against observational biases.

For $\Lambda_{\rm MSR}$ and $\Sigma_{\rm LDR}$, the individual stellar masses are likely accurate to within 20 -- 30\,per cent. This uncertainty on stellar mass in one sense is not important, as long as we correctly identify the subset of the ten most massive stars which are used to assess whether the region is mass segregated. Any stars hidden by differential extinction are likely to be lower mass, which would give the impression that the most massive stars are more centrally concentrated (i.e. $\Lambda_{\rm MSR} > 1$), but have lower than average surface densities ($\Sigma_{\rm LDR} < 1$). As we do not observe this combination, we can be resonably confident that our measurements are unaffected by this bias. As a check, we repeated our analyses excluding objects whose measured extinction exceeded $A_V > 4$ magnitudes and found no difference to our results.

A caveat associated with the simulations is that our $N$-body models assume instantaneous star formation, which is obviously an idealised situation. The observations of \citet{Foster15} show that star-forming cores seem to follow the gas structures on the outskirts of NGC\,1333. It is therefore possible that the central regions of NGC\,1333 (and indeed IC\,348) will acquire more pre-main sequence stars over time as filaments merge \citep[cf][]{Bate09,Myers11}. Our collapsing fractal simulations attempt to mimic this star-formation scenario, but fall short of modelling any of the gas physics. In this study $N$-body simulations do, however, hold the advantage in that they can be run to much older physical ages than hydrodynamical simulations, and can be directly compared to the observational data.

\subsection{Comparison with kinematic studies}

Recently, two separate studies have used high precision radial velocities to estimate the velocity dispersion of IC\,348 and NGC\,1333, and determined the dynamic state of each region by comparing the velocity dispersion with that expected if the region were in virial equilibrium using the virial mass estimate, $\sigma_{\rm vir}$, where 
\begin{equation}
\sigma_{\rm vir} = \sqrt{\frac{2GM}{\eta R}}.
\label{virial_mass}
\end{equation}
$M$ is the chosen mass enclosed within radius $R$, and $\eta$ depends on the density profile of the region in question, and is more difficult to determine.

\citet{Foster15} observed a velocity dispersion in NGC\,1333 of $\sigma = 0.92 \pm 0.12$\,km\,s$^{-1}$, compared to a virial mass estimate of $\sigma_{\rm vir} = 0.79 \pm 0.20$\,km\,s$^{-1}$, suggesting that the stars were in virial equilibrium. However, \citet{Foster15} also determined the velocity dispersion for the pre-stellar cores, and determined that they were subvirial. They hypothesised that the prestellar cores form with subvirial velocities, and two-body and violent relaxation causes the forming stars to speed up and attain virial equilibrium \citep{Parker16b}. 

Assuming that the prestellar cores observed \emph{today} have similar kinematic properties to the pre-stellar cores that formed the present-day pre-main sequence stars, our data are consistent with this collapsing scenario for NGC\,1333, especially given our conclusion that NGC\,1333 was likely to have been more dense in the past. 

In the case of IC\,348, \citet{Cottaar15} find it likely that this region is slightly supervirial, though still gravitationally bound. For very dense initial conditions ($\tilde{\rho} \sim 10^4$\,M$_\odot$\,pc$^{-3}$) the violent relaxation of initially subvirial clusters can create the illusion that they are supervirial and highly unbound following a cool-collapse \citep{Parker16b}. However, our analysis of the spatial diagnostics rules out such high initial stellar densities, suggesting that IC\,348 either formed without substructure (i.e.\,\,the present-day $\mathcal{Q}$-parameter reflects the initial conditions), or there may be an underlying problem with estimating the virial state from the velocity dispersion alone.

The scenario that IC\,348 formed without primordial substructure is unlikely; even modest dynamical evolution in a star-forming region with a stellar density of $\sim 10^2$M$_\odot$\,pc$^{-3}$ would result in the erasure of initial substructure \citep{Scally02,Parker12d,Parker13a,Parker14b,Parker14e} and the measured $\mathcal{Q} = 0.98$ would result from dynamical evolution of a medium density region (as shown in Fig.~\ref{ic348_nbody_med-a}). 

On the other hand, the use of the velocity dispersion \emph{in isolation} to determine the virial state of a region was shown to be problematic by \citet{Parker16b} for several reasons. First, the estimate of the velocity dispersion enclosed within the virial mass requires the determination of the structure parameter, $\eta$, which varies between $\sim 1 - 12$ depending on the amount of substructure present. Furthermore, the mass $M$ used to estimate the virial mass velocity dispersion in Equation~\ref{virial_mass} requires an estimate of the contribution of the gas potential, which is difficult to determine in IC\,348 \citep{Cottaar15}. Finally, the effects of binary orbital motion must first be removed from the velocity dispersion \citep{Cottaar12b}, which assumes the orbital properties of binaries in IC\,348 are similar to binaries in the Solar neighbourhood. It is far from clear that binaries do have universal orbital properties in nearby star-forming regions \citep{King12b,Parker14e}.

\subsection{A typical density for star formation?}

It is worth placing the derived initial densities ($\tilde{\rho} \sim 100 - 500$\,M$_\odot$\,pc$^{-3}$ for IC\,348 and $\tilde{\rho} \sim 500 - 2000$\,M$_\odot$\,pc$^{-3}$ for NGC\,1333) in the context of those in other nearby star-forming regions. In Taurus, $\rho$~Ophiuchus and Chamaeleon~I, the $\mathcal{Q} - \Sigma_{\rm LDR}$ diagram suggests initial densities very similar to their observed present-day density \citep{Parker14e,Sacco17}. All three regions are low-mass (as are IC\,348 and NGC\,1333 and many other star-forming regions within 1\,kpc of the Sun). 

The Orion Nebula Cluster is the most massive nearby star-forming region, and its position on the  $\mathcal{Q} - \Sigma_{\rm LDR}$ diagram points towards high-density ($\tilde{\rho} > 10^3$\,M$_\odot$\,pc$^{-3}$) initial conditions. However, spatial and kinematic analysis of the more distant but massive Cyg~OB2 association is also consistent with low-density initial conditions \citep{Wright14,Wright16}. A larger sample of star-forming regions should be analysed, but our results in tandem with others suggest that the typical stellar density of star formation is likely to be around  $\tilde{\rho} \sim 10^2 - 10^3$M$_\odot$\,pc$^{-3}$.

Interestingly, these stellar densities straddle the border between the density regime ($ \tilde{\rho} >$1000\,M$_\odot$\,pc$^{-3}$)  where dynamical interactions could significantly alter planetary systems and discs \citep{Adams06,Parker12a,Vincke15} and affect the orbital properties of binary systems \citep{Kroupa95a,Parker12b} and much more benign initial conditions ($\tilde{\rho} \sim <$100\,M$_\odot$\,pc$^{-3}$) where dynamical interactions are unimportant. Clearly, a larger census of star-forming regions is required to address whether there exists a typical density for star formation, and kinematic data from \emph{Gaia} \citep{Brown16} and associated spectroscopic surveys such as Gaia-ESO \citep{Gilmore12} will be crucial in our understanding of this problem. 

\section{Conclusions}
\label{conclude}

We have presented an analysis of the spatial distributions of stars in two nearby star-forming regions in the Perseus molecular cloud, the 3\,Myr old IC\,348 and the 1\,Myr old NGC\,1333. We determine the structure as measured by the $\mathcal{Q}$-parameter, the $\Lambda_{\rm MSR}$ measure of mass segregation and the $\Sigma_{\rm LDR}$ measure of the relative surface densities of the most massive stars compared to low-mass stars. We then compare these values to $N$-body simulations of the evolution of star-forming regions with a range of initial conditions. Our conclusions are the following.

(i) IC\,348 has a $\mathcal{Q}$-parameter of 0.98, $\Lambda_{\rm MSR} = 1.1^{+0.2}_{-0.3}$ and $\Sigma_{\rm LDR} = 1.77$. Compared to $N$-body simulations at similar ages (2 -- 3\,Myr), these values suggest subvirial, medium density ($\tilde{\rho} \sim 10^2$M$_\odot$\,pc$^{-3}$) initial conditions. 

(ii) Similarly, NGC\,1333 has a $\mathcal{Q}$-parameter of 0.89, $\Lambda_{\rm MSR} = 1.2^{+0.4}_{-0.3}$ and $\Sigma_{\rm LDR} = 1.03$. Compared to $N$-body simulations at similar ages ($\sim$1\,Myr), these values also suggest subvirial velocities, but with medium-to-high density ($\tilde{\rho} \sim 10^3$M$_\odot$\,pc$^{-3}$) initial conditions. 

(iii) The spatial analysis of NGC\,1333 is consistent with estimates of its virial state using the radial velocity dispersion, whereas IC\,348 is slightly discrepant in that the spatial analysis suggests the region is in virial equilibrium, whereas the velocity dispersion hints at a supervirial state. 

(iv) The initial inferred density for IC\,348 is similar to its present-day density ($\tilde{\rho} \sim 10^2$M$_\odot$\,pc$^{-3}$), but is higher for NGC\,1333. In comparison to other nearby star-forming regions, these initial densities are lower than that inferred for e.g.\,\,the Orion Nebula Cluster, but higher than the nearby Taurus star-forming region.\\

Based on these results, and other nearby star-forming regions \citep{Parker14e}, the typical initial stellar density for star-forming regions is between $\tilde{\rho} \sim 10^2 - 10^3$\,M$_\odot$\,pc$^{-3}$, which is dense enough to perturb a moderate fraction \citep[up to $\sim 10$\,per cent;][]{Parker12a} of protoplanetary discs and binary stars.

\section*{Acknowledgments}

We thank the anonymous referee for their comments and suggestions, which have greatly improved the paper. We also thank Kevin Luhman for making his data available to us prior to publication, and for comments and suggestions on the original manuscript. We thank the authors of \citet{Parker14b} for providing the material for Section 3.1. RJP acknowledges support from the Royal Society in the form of a Dorothy Hodgkin Fellowship.

\bibliographystyle{mn2e}
\bibliography{general_ref}

\appendix

\section{$N$-body simulations with a statistical IMF}
\label{appendix}

The form of the initial mass function (IMF) is generally universal, at least on local scales \citep{Bastian10,Offner14}. Both IC\,348 and NGC\,1333 are consistent with the common analytical approximations for the form of the IMF in the literature \citep[e.g.][]{Kroupa93,Chabrier05,Maschberger13}. However, due to the low numbers of stars in each region (425 for IC\,348 and 150 for NGC\,1333), neither region has a fully sampled IMF and statistical variations can lead to noticeably different IMFs if they are randomly sampled \citep{Elmegreen06,Parker07,Maschberger08}.

  \citet{Kouwenhoven14} show that subtle changes to the IMF of a star cluster can lead to differences in the properties of the clusters at later ages and in this Section we examine the evolution of an IC\,348-like star-forming region with masses drawn from a \citet{Maschberger13} IMF instead of using masses from the \citet{Luhman16} datasets. We repeat the simulation from Section~\ref{nbody_comp} with 425 stars and an initial radius of $r_F = 1.5$\,pc and draw masses from the probability distribution function in \citet{Maschberger13}:
  \begin{equation}
    p(m) \propto \left(\frac{m}{\mu}\right)^{-\alpha}\left(1 + \left(\frac{m}{\mu}\right)^{1 - \alpha}\right)^{-\beta}
    \label{imf}.
  \end{equation}
  Here, $\mu = 0.2$\,M$_\odot$ is the average stellar mass, $\alpha = 2.3$ is the \citet{Salpeter55} power-law exponent for higher mass stars, and $\beta = 1.4$ is used to describe the slope of the IMF for low-mass objects \citep*[which also deviates from the log-normal form;][]{Bastian10}. Finally, we sample from this IMF within the mass range $m_{\rm low} = 0.01$\,M$_\odot$ to $m_{\rm up} = 50$\,M$_\odot$.
  
\begin{figure*}
  \begin{center}
    \setlength{\subfigcapskip}{10pt}
    \hspace*{-1.3cm}\subfigure[$\mathcal{Q}$-parameter]{\label{ic348_nbody_masch-a}\rotatebox{270}{\includegraphics[scale=0.29]{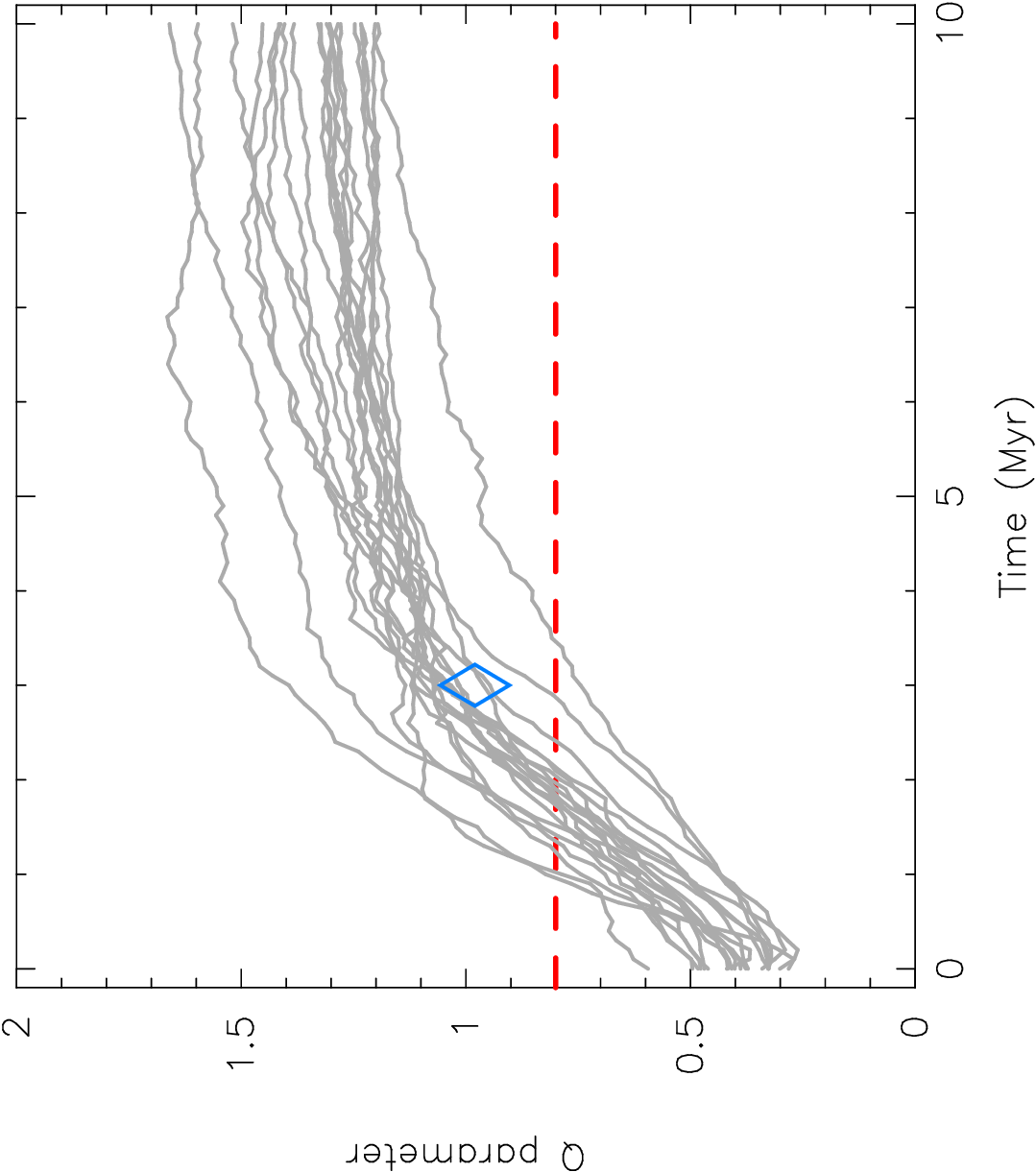}}}
    \hspace*{0.3cm}
    \subfigure[$\mathcal{Q} - \Lambda_{\rm MSR}$]{\label{ic348_nbody_masch-b}\rotatebox{270}{\includegraphics[scale=0.29]{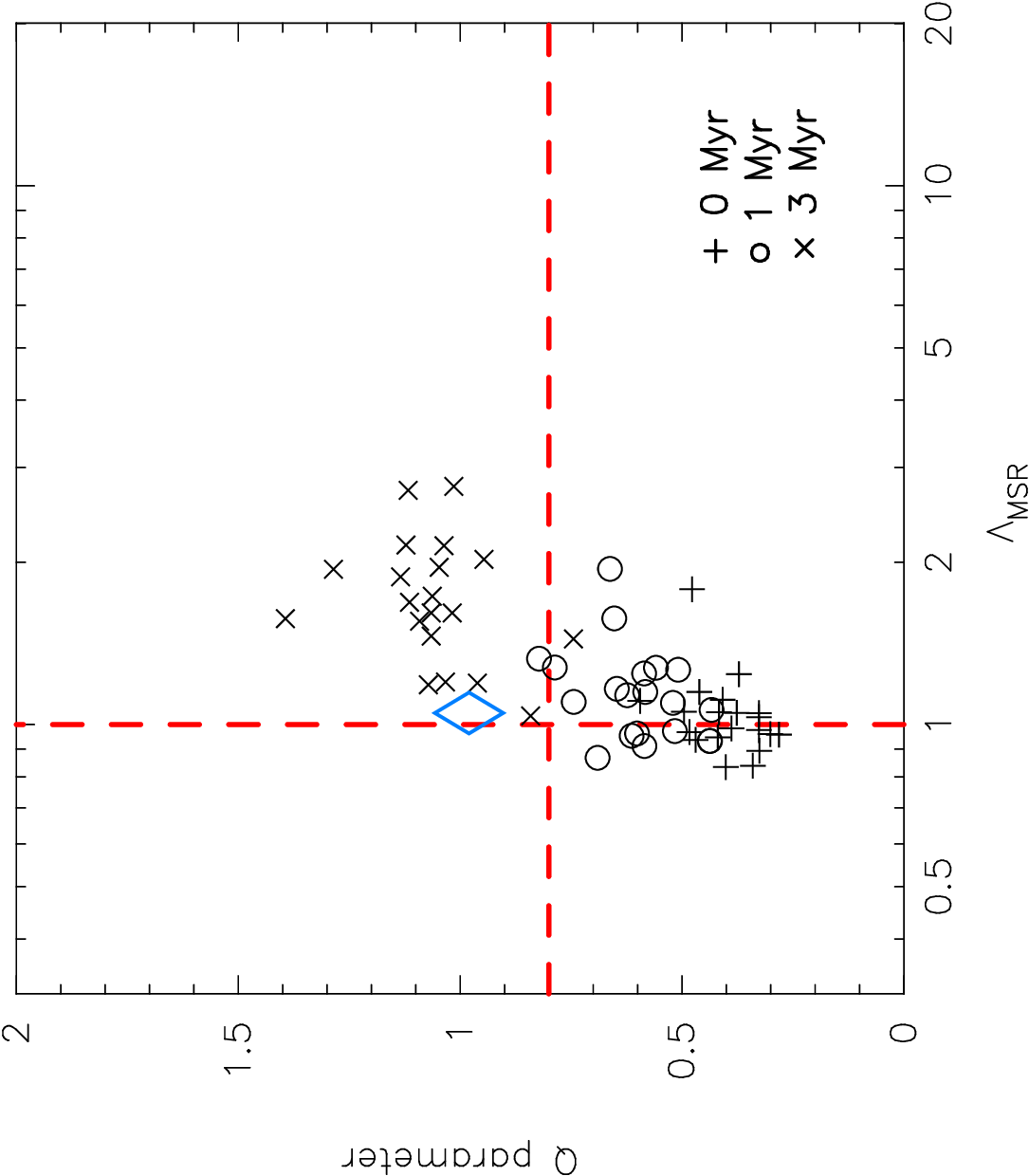}}}
    \hspace*{0.3cm}\subfigure[$\mathcal{Q} - \Sigma_{\rm LDR}$]{\label{ic348_nbody_masch-c}\rotatebox{270}{\includegraphics[scale=0.29]{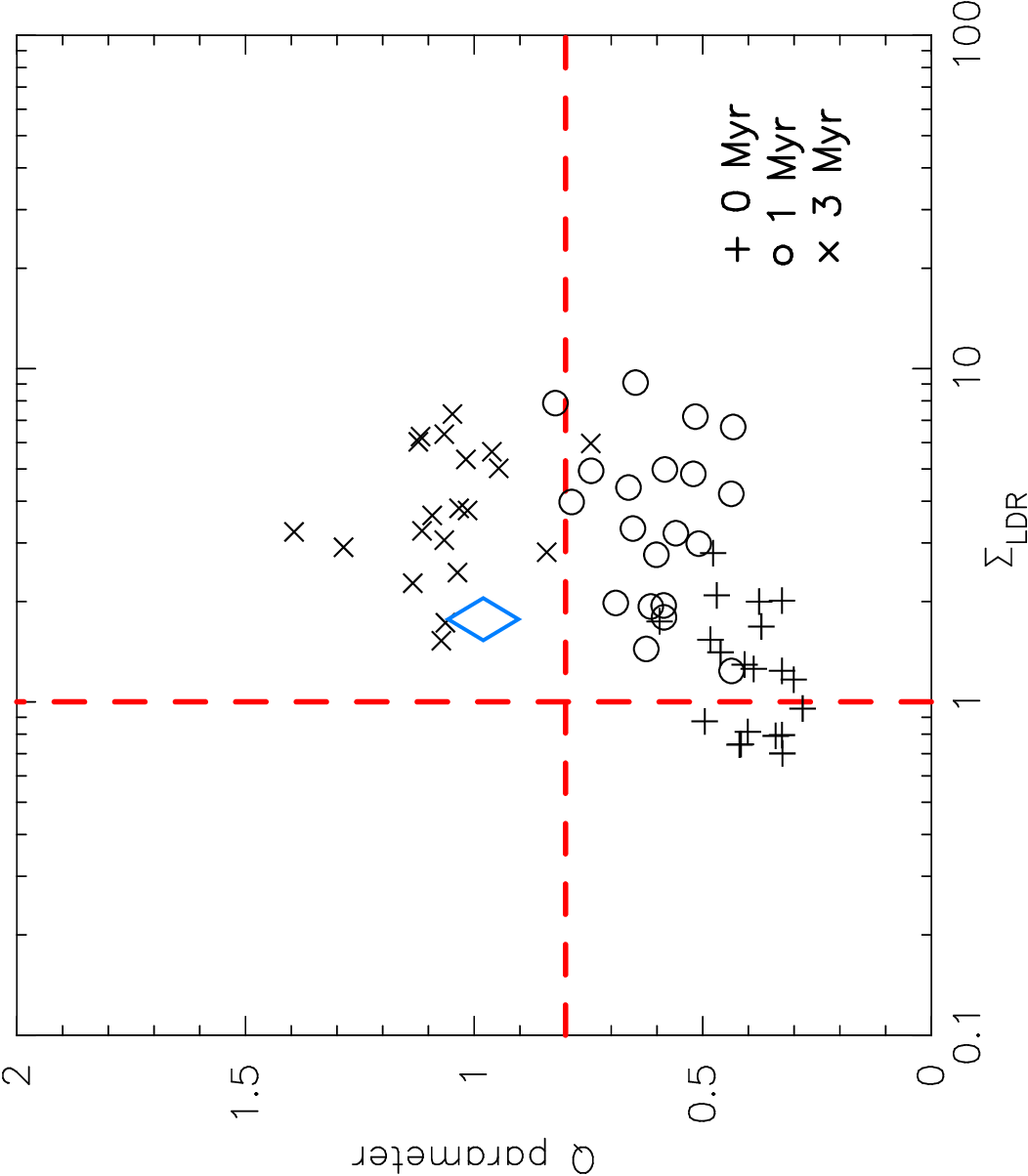}}}
    \caption[bf]{As Fig.~\ref{ic348_nbody_med} but with stellar masses drawn randomly from an initial mass function. $N$-body simulations of star-forming regions containing 425 stars undergoing cool-collapse with stellar masses drawn from a \citet{Maschberger13} IMF and initial stellar densities of 100 -- 500\,M$_\odot$\,pc$^{-3}$. In panel (a) we show the evolution of the $\mathcal{Q}$-parameter; in panel (b) we show the $\mathcal{Q}-\Lambda_{\rm MSR}$ plot and in panel (c) we show the $\mathcal{Q}-\Sigma_{\rm LDR}$ plot. in panels (b) and (c) the simulation data at 0, 1 and 3\,Myr are indicated by the plus signs, open circles and crosses, respectively. In all three panels, the blue diamond symbol shows the observational value at 3\,Myr. }
    \label{ic348_nbody_masch}
  \end{center}
\end{figure*}

The evolution of this star-forming region is shown in Fig.~\ref{ic348_nbody_masch}, which should be directly compared to Fig.~\ref{ic348_nbody_med}. We also performed a similar experiment for an NGC\,1333-like cluster, instead drawing 150 stars from Eqn.~\ref{imf} and comparing with the evolution of the simulations in Fig.~\ref{ngc1333_nbody}. The variation between the simulations for both regions is not beyond what we would expect in the stochastic evolution of low-mass systems \citep{Allison10,Parker14b} and we conclude that the mass function measured by \citet{Luhman16} does not result in a different dynamical history from a similar region with a field-like IMF.

\label{lastpage}

\end{document}